\documentclass[aps,twocolumn,prb,groupedaddress]{revtex4-2}

\usepackage{graphicx}
\usepackage{amsmath,amssymb}
\usepackage{times}
\usepackage[colorlinks,citecolor=blue]{hyperref}
\usepackage{tcolorbox}
\usepackage{caption}
\usepackage{subcaption}
\usepackage{soul}
\setstcolor{red} 
\usepackage{float}

\newcommand{\revisiontwo}[1]{{\color{black} #1}}

\usepackage{xr}
\begin{document} 
\title{Electride Behavior in High-Pressure Silicon and Other Elements in Solid and Liquid Phases}

\author{Salma Ahmed}
\email{salma\_ahmed@berkeley.edu}
\affiliation{Department of Earth and Planetary Science, University of California, Berkeley, California 94720, USA}
\author{Felipe Gonz\'alez-Cataldo}
\email{f\_gonzalez@berkeley.edu}
\affiliation{Department of Earth and Planetary Science, University of California, Berkeley, California 94720, USA}
\author{Victor Naden Robinson}
\email{victornadenrobinson@gmail.com}
\affiliation{Department of Earth and Planetary Science, University of California, Berkeley, California 94720, USA}
\author{Burkhard Militzer}
\email{militzer@berkeley.edu}
\affiliation{Department of Earth and Planetary Science, University of California, Berkeley, California 94720, USA}
\affiliation{Department of Astronomy, University of California, Berkeley, California, USA}

\date{\today}

\begin{abstract}
Electrides are materials in which some of the electrons are localized at the interstitial sites rather than around the atoms or along atomic bonds. Most elemental electrides are either alkali metals or alkaline-earth metals because of their low ionization potential. In this work, we report that elemental silicon becomes an electride at pressures exceeding 400 GPa. With {\it ab initio} molecular dynamics (MD) simulations, we study this behavior for silicon, sodium, potassium, and magnesium at high pressure and temperature. We performed simulations for liquids and ten crystal structures. Charge density and electron localization functions (ELF) are analyzed for
representative configurations extracted from
the MD trajectories. By analyzing a variety of electride structures, we suggest the following quantitative thresholds for the ELF and charge density in each interstitial site to classify high-pressure electrides: (1) the maximum ELF value should be greater than 0.7, (2) there should be at least 0.9 electrons near the ELF basin, and (3) the Laplacian charge density, $\nabla^2 \rho(\mathbf{r}_0)$, should be negative with magnitude greater than $10^{-3}\ e/\mathrm{bohr}^5$. Finally, we compute X-ray diffraction patterns to determine the degree to which they are affected by the electride formation. 
Overall, this framework could become a benchmark for future theoretical and experimental studies on electrides.
\end{abstract}

\maketitle

\section{Introduction}

Electrides are a fascinating class of materials in which some valence electrons are relegated to interstitial sites rather than forming bonds or being associated with atoms. In most materials, the valence electrons form ionic, covalent, or metallic bonds. Electrides, however, are compounds in which the electrons are localized away from the nuclei, similar to those of anions.
\revisiontwo{One of the first electrides was discovered by Dawes \emph{et al.} in 1986 with Cs$^+$(18C6)$_2$e$^-$~\cite{dawes_first_1986}. Similar organo-metallic salts such as Li$^+$(cryptand[2.2.2])Cs$^-$ and Cs$^+$(cryptand[2.2.2])Cs$^-$ have also been reported to have electride behavior at ambient pressures due to the organic anion zigzag structure ~\cite{ichimura_one-dimensional_2006}. These ambient-pressure electrides have been known to have electron-rich regions within the crystal with 1-2 electrons that exceed the traditional molecular or crystal stoichiometry. 
As a result, these excess electrons are delocalized in between molecules in 2D sheets~\cite{dawes_first_1986,ichimura_one-dimensional_2006, Wan_BaCu_2024} or 3D molecular cages~\cite{C12_matsuishi_2003,C12_Li_2004} or within the interstitial sites of a crystal~\cite{chanhom_sr3crn3_2019,Li_Ca5Pb3_2021}.}

\revisiontwo{Another} class of electrides, \revisiontwo{which will be the focus of this article,} is high-pressure electrides. \revisiontwo{Pressures typically} in the GPa range compress the electron orbitals, causing their orbital energies to increase. The orbital energies then become so high that the electrons in the highest-energy orbitals (typically, the s and p orbitals~\cite{miao_high_2014}) occupy the interstitial sites of the solid. 
Previous \revisiontwo{high-pressure experiments} have focused on the electride behavior in alkali metal and alkaline earth metal solids~\cite{ma_transparent_2009,woolman_structural_2018,yu_optical_2018,gorman_experimental_2022,chanhom_sr3crn3_2019,racioppi_Ca_electride_2025} because their low ionization energies \revisiontwo{lead to electride transitions} at lower pressures~\cite{miao_high_2014} \revisiontwo{that can be achieved using diamond-anvil cells}. \revisiontwo{Unlike ambient-pressure electrides, there is no stoichiometric change to the material when it reaches the electride state. In addition, high-pressure electrides tend to have their expelled electrons be more localized than their ambient-pressure counterparts.} Computational studies have shown electride behavior for simple metal solids~\cite{miao_high_2014} as well as in more binary and ternary inorganic systems~\cite{zhou_LiFe_2016,zhou_NaFe_2016,zhang_YSi_2021,zhang_inorganic_elf_2017}. A common trend with most previous electride work is that they focus on one solid structure at either 0~K (in theory) or room temperature (in experiment), but very few studies have looked at electride liquids~\cite{zong_free_2021,qin_electride_2025,gonzalez-cataldo_structural_2023} or high-temperature solids~\cite{paul_thermal_2020}. Instead, there is a stronger emphasis on the relationship between electride behavior and pressure or chemical formula, but not on temperature or structure. 

Electride formation is often associated with low ionization potential~\cite{miao_high_2014}, hence why most electrides include alkali metals and/or alkaline-earth metals. Miao and Hoffmann introduced a He-containment model to predict the likelihood of electride behavior for said metal solids up to 500 GPa~\cite{miao_high_2014}. This model states that the interstitial site of the lattice is analogous to a vacant 1s orbital (which is filled in a He atom). According to that model, carbon does not favor the electride state at 500 GPa, but Martinez-Canales \emph{et al.} have found electride behavior for solid carbon in the fcc configuration at 21 TPa~\cite{martinez-canales_thermodynamically_2012}. This brings the question: how many more compounds can be electrides if we expand on the pressure we study?  

In this work, we show that elemental silicon can also be an electride. 
While experimental efforts have investigated high-pressure silicon structures~\cite{gong_x-ray_2023,lin_temperature-_2020}, silicon is traditionally considered unlikely to exhibit electride properties due to its high ionization energy. 
This highlights both the novelty and potential challenges of exploring the behavior of silicon under extreme conditions, where it might exhibit electride properties. Silicon assumes the diamond crystalline structure at ambient pressure, which transforms into the fcc structure at 80~GPa and to the bcc structure around 3000~GPa~\cite{Paul_silicon_2019}.
We characterize the electride behavior of Si and compare it to three known electride elements (Mg, K, and Na) 
in the solid and liquid states. We establish criteria to distinguish an electride from a non-electride phase at high pressure based on the amount of electronic charge that is enclosed in the ``pockets" in interstitial sites. We compare the geometry of these pockets between different elements/structures at the ground state, and also discuss how temperature affects electride behavior. By investigating a variety of electride elements and structures, we can better define a quantitative criterion for electride behavior and open doors to future studies to find electrides in other elements at higher pressures. 

\section{Methods}

We conducted density functional theory (DFT) calculations using the Vienna \emph{ab initio} Simulation Package (VASP)~\cite{VASP}, employing the PBE functional~\cite{PBE} and PAW pseudo-potentials~\cite{PAW} with 12 valence electrons for Si, 7 for Na and K, and 10 for Mg. For Si, we examined the fcc and bcc structures, performing electronic structure calculations across a range of pressures with a dense sampling of the Brillouin zone for our unit cells (see Table~S1). Similarly, for Mg, we considered the fcc, bcc, sc, and sh structures, while for K, we investigated the bcc, fcc, and \emph{tI}19 structures. For Na, we explored the \emph{cI}16 and h\textit{P}4 structures.

\subsection{Studying electride behavior with respect to temperature}
We generated supercells for each structure by replicating the respective unit cells. The number of atoms in each of the corresponding supercells are indicated in Table~S1.
For the MD simulations, we used $\Gamma$-point sampling for the Brillouin zone of all super cells. The structures were thermalized in the NVT ensemble using the Nos\'e-Hoover thermostat~\cite{nose_unified_1984,hoover_canonical_1985} for at least 2 ps to ensure convergence of thermodynamic quantities like energy and pressure. 

To determine the effect of temperature on the electride behavior, we simulated Si, Na, K, and Mg using DFT at three different temperatures: $T=T_0 = 0$~K (unit cell calculations), $T_1$ with $T_0 < T_1 < T_m$ (solid at high temperature), and at $T_2 > T_m$ (liquid) where $T_m$ is the melting temperature for each material.
Table~\ref{tab:MD_table} lists the structures of each material that we simulated in pressure ranges where they have been reported to be stable. 
We chose these structures to compare the electride behavior of Si to known electrides. Of the structures in Table~\ref{tab:MD_table}, the known electrides we simulated were Na \emph{cI}16~\cite{woolman_structural_2018} and hP4~\cite{ma_transparent_2009,racioppi_electride_2023}, Mg fcc~\cite{Li_Mg_electride_2010}, sh~\cite{Li_Mg_electride_2010} and sc~\cite{gonzalez-cataldo_structural_2023,gorman_experimental_2022}, and K \emph{tI}19~\cite{woolman_structural_2018}. The incommensurate Na \emph{cI}16 and K \emph{tI}19 structures are host-guest structures that have a large host lattice and guest atoms that form a chain within it (Fig.~S2). To make sure that our proposed criterion distinguishes an electride from a non-electride, we simulated regular metals K bcc and fcc, and Mg bcc.


For the heated simulations (both solid and liquid phases), we selected 10--20 equally spaced snapshots from each MD trajectory using the blocking method ~\cite{Flyvbjerg1989} shown in Supplementary Information ~\cite{SuppMat}. In the supplementary information (Fig.~S7), we show that the ELF and charge density statistics do not change when we include additional snapshots. In each snapshot, we identified all non-nuclear critical points (i.e., maxima in the ELF that are not located at atomic nuclei) and evaluated related quantities such as the value of ELF at the maximum and the integrated charge per pocket. \revisiontwo{We only evaluated the non-nuclear attractors with ELF local maxima greater than 0.7 and excluded all others when we computed mean values of the related quantities by averaging over all critical points in the selected snapshots.}
\revisiontwo{These final mean values were compared with critical values to determine whether the material is an electride.}
In liquids, the number, position, and shape of charge pockets vary between snapshots, leading to larger fluctuations in these quantities compared to the solid phases. By averaging over both spatial and temporal variations, we obtain representative mean values and associated standard deviations, which capture the inherent disorder in the liquid state while allowing meaningful comparison between systems.
We verified that increasing the number of snapshots did not change the ELF and charge density statistics (see Supplementary Fig.~S7), confirming the robustness of our results.
Finally, we revisit the electride properties of Mg, Na, and K, which have previously been reported to be electrides in either the solid or the liquid phase~\cite{zong_free_2021,gonzalez-cataldo_structural_2023,paul_thermal_2020}.

\begin{table}[h!]
    \centering
    \begin{tabular}{c c c c c }\hline
        Element & Pressures (GPa) & Structure & Solid T (K) & Liquid T (K) \\\hline
         Si & 200 - 3,000 & fcc & 1,000 & 35,000 \\
         Si & 3,200 - 3,700 & bcc & 2,000 & 35,000 \\
         Na & 150 - 200 & \emph{cI}16 & 300 & 3,000 \\
         Na & 210 - 300 & hP4 & 500 & 3,000\\
         K & 1 - 10 & bcc & 200 & 3,000 \\
         K & 11 - 20 & fcc & 400 & 3,000 \\
         K & 21 - 30 & \emph{tI}19 & 300 & 3,000 \\ 
         Mg & 50 - 450 & bcc & 4,000 & 20,000\\
         Mg & 450 - 750 & fcc & 4,000 & 20,000\\
         Mg & 750 - 1000 & sh & 4,000 & 20,000\\
         Mg & 1000 - 1400 & sc & 4,000 & 20,000\\\hline
    \end{tabular}
    \caption{List of elements, pressure ranges, structures, and the temperatures set to heat the solids and liquids. Adapted from~\cite{Paul_silicon_2019, ma_transparent_2009,racioppi_electride_2023, Li_Mg_electride_2010, gonzalez-cataldo_structural_2023,gorman_experimental_2022}.
    }
    \label{tab:MD_table}
\end{table}

\subsection{Topological analysis of the ELF and charge density}

\newcommand{\rr}{\mathbf{r}}

Understanding the topology of the charge density is key to classifying electrides and quantifying the number of localized, interstitial electrons. 
Although there is a widely accepted conceptual definition of an electride, there is no universally adopted set of quantitative criteria for identifying and characterizing this behavior.
The metric of localization and anion-like behavior has varied from study to study, but the electron localization function (ELF) has been a common metric to determine electron localization. In this paper, we will refer to the interstitial sites of high localization as \textit{charge pockets}.
While previous electride studies have called for high ELF values as a means to identify electride behavior, it is unclear what the minimum value needed is.
Zhang~\emph{et al.}~\cite{zhang_inorganic_elf_2017} developed a ``computer-assisted inverse-design" method to search for inorganic electrides. In their study, they propose that ELF values above 0.75 are indicators of proper electron localization. In this work, we will discuss how ELF values vary between elements and structures, but ultimately, having an ELF value greater than 0.7 is a valid starting point for electride classification. Having an ELF value greater than 0.7 and above ensures a local maximum in the charge density, and having both the ELF and charge density is needed to classify a material as an electride. 
For both the charge density and ELF, the local maxima and minima represent different parts of the crystal, such as atoms and bonds. Since electrides have electrons localized away from the atoms, finding local maxima in ELF and charge density that are not attributed to atoms or bonds is necessary. Additionally, most materials metalize at high pressures, so it is important when studying electrides that exceed metallization pressures to distinguish localized charge pockets from delocalized electrons. \revisiontwo{This differs from ambient-pressure electrides, where the expelled electrons are delocalized due to the electron-rich nature of the material. Because of this delocalized behavior, ambient-pressure electrides can exhibit lower ELF values (0.5 and lower)~\cite{C12_matsuishi_2003,C12_Li_2004}. Due to the different mechanisms that create ambient-pressure and high-pressure electrides, our ELF threshold of 0.7 is only appropriate for classifying high-pressure electrides.}

The electron localization function (ELF)~\cite{Becke1990} provides a measure of the degree of electron localization and is defined by,
\begin{equation}
    {\rm ELF(\rr)}=\frac{1}{1+\left(D_\sigma(\rr)/D_\sigma^0(\rr\right))^2},
\end{equation}
where $D_\sigma(\rr)$ is the leading term of the Taylor expansion of the spherically averaged probability of finding a $\sigma$-spin electron around a reference $\sigma$-spin electron located at $\rr$. For a given spin state, $\sigma$, $D_\sigma(\rr)$ is derived from the kinetic energy density, $\tau_\sigma(\rr)$, and the electron \textit{spin} density, $\rho_\sigma(\rr)$, such that, 
\begin{equation}
    D_\sigma(\textbf{r}) = \tau_\sigma(\mathbf{r}) - \frac{1}{4} \frac{(\nabla\rho_\sigma(\mathbf{r}))^2}{\rho_\sigma (\mathbf{r})}\;\;.
\end{equation}
where $\tau_\sigma$ is the bosonic kinetic energy density and $\rho_\sigma$ is the electron-spin density (different from $\rho$, which is the charge density).
To introduce a normalization, the ELF includes the term $D_\sigma^0=\frac{3}{5} (6\pi^2)^{2/3}\rho_\sigma^{5/3}$, which corresponds to a uniform electron gas with spin-density equal to the local value of $\rho_\sigma(\rr)$. For $D_\sigma=D_\sigma^0$, one obtains an ELF value of 0.5. 
Values higher than 0.5 indicate enhanced localization compared to a free-electron (metallic) behavior. In summary, the ELF provides a scale that ranges from 0 (completely delocalized) to 1 (perfect localization), or more accurately, from 0.5 (delocalized) to 1 (perfect localization)~\cite{savin_elf_1997}. 

With that definition, covalent bonds require pairs of electrons to occupy localized orbitals and will thus lead to high ELF values. The ELF is a useful metric for studying electron localization, but by itself, it is insufficient to identify electride behavior, which requires electrons to be localized in charge basins at interstitial sites. The ELF, however, also assumes high values at the nuclei because of the occupation of atomic orbitals, which we need to exclude when we identify electride behavior. Furthermore, the ELF is not a direct measurement of the electron density. So, once an interstitial charge basis has been identified from the ELF, we revert to charge density to measure how much charge it contains.

We use the code \verb+critic2+~\cite{gatti_chemical_2005,otero-de-la-roza_finding_2022} to find all of the non-nuclear local maxima in the ELF and charge density landscape and determine their volume and shape. 
\verb+Critic2+ does this by analyzing the gradient of either the charge density or the ELF. It finds all critical points (i.e., where the gradient is zero) and classifies each critical point by taking the sum of the signs of the gradient path between two points. 
Conversely, if all 3 directions are positive, the critical point is a local minimum. Other combinations where some directions have positive and negative gradient paths can imply bond paths and ring surfaces~\cite{otero-de-la-roza_finding_2022}. The local maxima associated with atoms or charge pockets are surrounded by regions of steep gradients where the charge density varies rapidly,  as captured by the large negative values of its Laplacian, $\nabla^2 \rho$. 

This topological analysis can also be applied to the ELF. As discussed in previous studies~\cite{Postils_molecular_electride_2015,racioppi_electride_2023, Racioppi2024}, an electride needs to have local maxima in both the ELF and charge density at lattice sites with no atoms. In this study, we integrate the charge density around the ELF local maxima (also referred to as ELF basins). The limits of integration are determined by the zero-flux surface of the ELF or charge density gradient, initially proposed by Bader~\cite{bader_atoms_1985} and implemented in DFT by Henkelman \emph{et al.}~\cite{henkelman_fast_2006}. This method is similar to BadELF~\cite{weaver_counting_2023}, except the zero-flux surfaces are used to partition the atoms as well (as opposed to Voronoi-tessellation). We utilize the weighted method proposed by Yu and Trinkle~\cite{yu_accurate_2011} for Bader volume integration. This method reduces the error of having a discrete grid by weighting the grid units based on how much the boundary crosses them.
In summary, we identify ELF basins using the Bader analysis in \verb+critic2+ and then integrate the charge density inside these basins using the Yu-Trinkle algorithm~\cite{yu_accurate_2011}
to determine the number of electrons per pocket.

\subparagraph{Quantitative criteria for electride behavior.} While there is no consensus yet on how to identify electrides, three criteria are generally proposed by Postils et al. ~\cite{Postils_molecular_electride_2015}: the presence of a non-nuclear attractor of electron charge, a negative value of the Laplacian of the electron density at this attractor, and the presence of an ELF maximum. In this work, we propose the following unified and quantitative set of required conditions for a material and phase to be an electride: 

\begin{enumerate}
\item \textbf{Presence of charge basins:}  
There must be a local maximum of charge density that does not coincide with an atomic position or a covalent bond. These local maxima are referred to as \textit{charge basins}.
\item {\bf Presence of ELF maxima}: In the vicinity of the charge density maximum, there must be a non-nuclear local maximum in the ELF where its value exceeds ELF$^*$=0.7. 
\item {\bf Charge density integration}: \revisiontwo{We require average charge per basin} to be at least $c^*=0.9$ electrons, for which we integrate charge density inside the boundary of every basin with an ELF maximum greater than 0.7. As a boundary, we follow Otero-de-la-Roza~\cite{otero-de-la-roza_finding_2022} and adopt the no-flux boundary conditions of the ELF, $\mathbf{n} \cdot \nabla \mathrm{ELF}(\rr) =0$, as implemented in the \verb+critic2+ code. $\mathbf{n}$ is the normal vector of the boundary. 
\item {\bf Large, negative Laplacian of the charge density}: At the charge density maxima, $\rr_0$, \revisiontwo{we require the average value} of the Laplacian to be at least $-\nabla^2 \rho(\mathbf{r}_0) \geq L^* = 10^{-3}\ e/\mathrm{bohr}^5$ so that the curvature at the interstitial site is larger than values of a typical bond. 
\end{enumerate}

Our high-pressure electride criteria rely on three critical values: ELF$^*=0.7$, $c^*=0.9$, and $L^*=10^{-3}\ e/\mathrm{bohr}^5$. We have chosen these plausible values so that they work for different materials and solid and liquid phases, while previously identified electrides still satisfy them. Not surprisingly, if one adopted higher critical values, fewer materials would be considered to be 
electrides. 
For example, if ELF$^*$ were increased to 0.8, it would exclude ambient-pressure inorganic electrides~\cite{zhang_inorganic_elf_2017}.
If one increased $c^*$ to 1.0, Mg would no longer be an electride in the simple hexagonal (sh) phase (Table.~S3). 
 


When applying this criterion, we began by identifying all critical points of the ELF (points where the gradient is zero). Critical points located at the nuclei were excluded, as were those corresponding to chemical bonds (as classified \texttt{critic2}). From the remaining interstitial ELF maxima, we retained those with $\mathrm{ELF}_{\mathrm{max}} > 0.7$.
For each of these candidates, we verified that the charge density ($\rho$) also has a local maximum at the same position, ensuring that the ELF feature corresponds to a genuine electron density concentration. 
We then integrated the charge density within the corresponding basin to ensure that each pocket contained at least 0.9 electrons.
Finally, we calculated the Laplacian of the charge density at the maximum, discarding any point whose value did not exceed the critical threshold $L^*$. This procedure ensures that only physically meaningful, well-localized interstitial electron pockets are counted.

\section{Results}\label{sec:results}

\subsection{Electride Behavior at $\mathbf {T = 0}$ K}
\subsubsection{Silicon}
\begin{figure}[!hbt]
\centering 
\includegraphics[width=3.5cm]{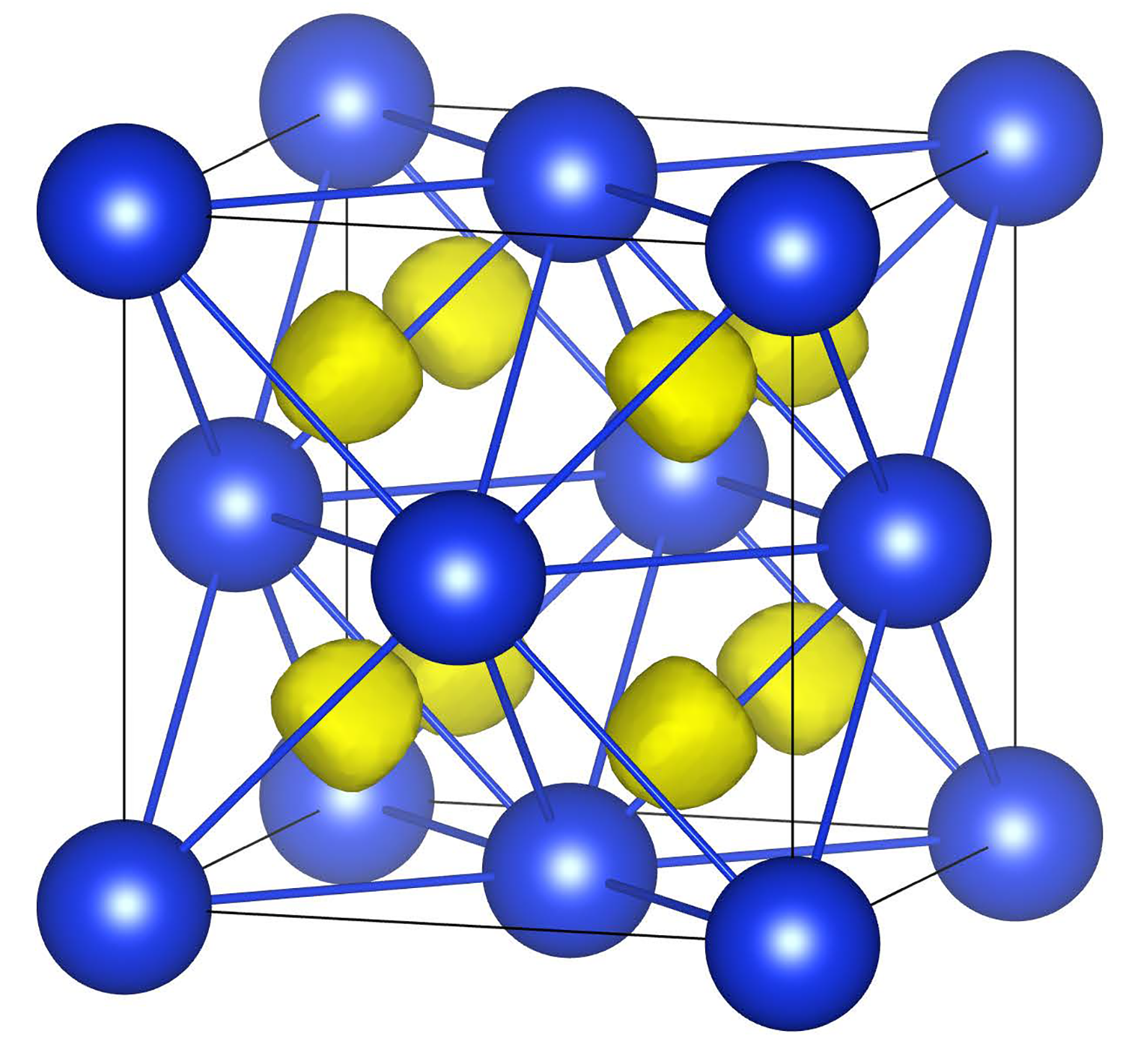}
\includegraphics[width=3.5cm]{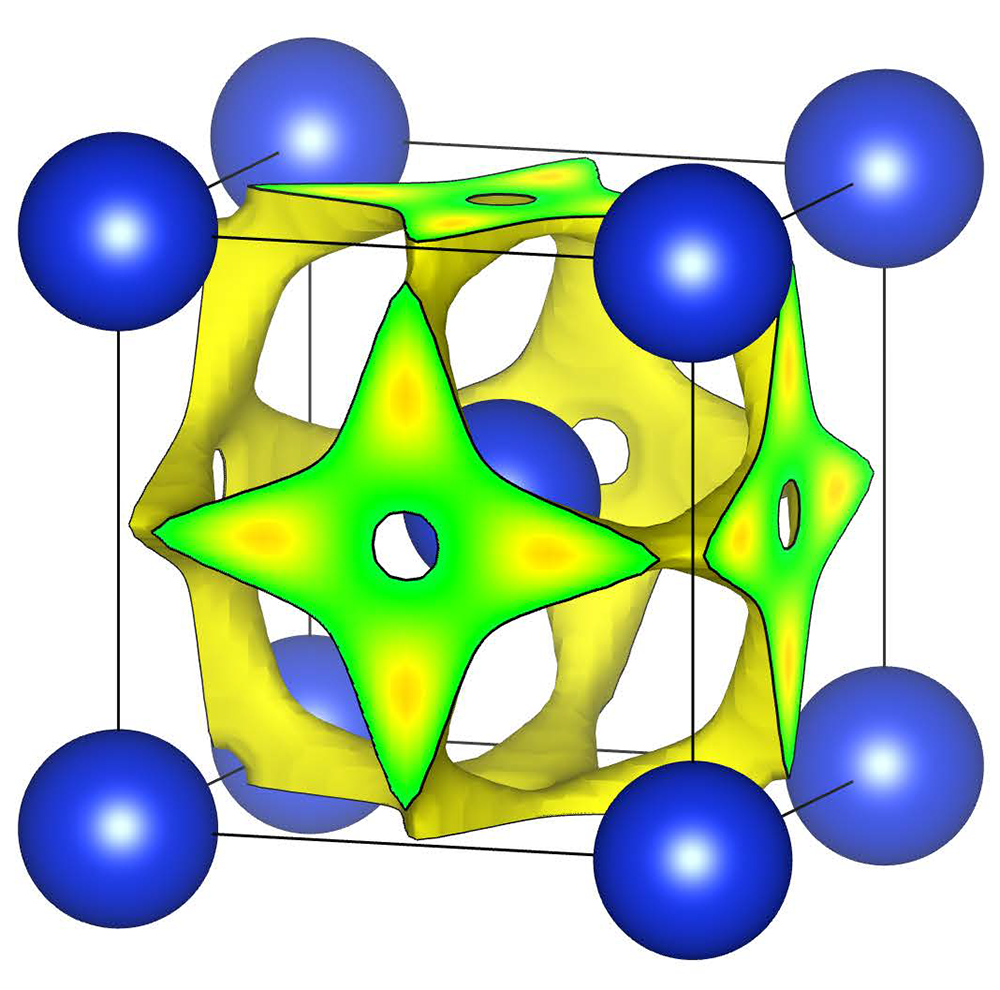}
\includegraphics[width=3.5cm]{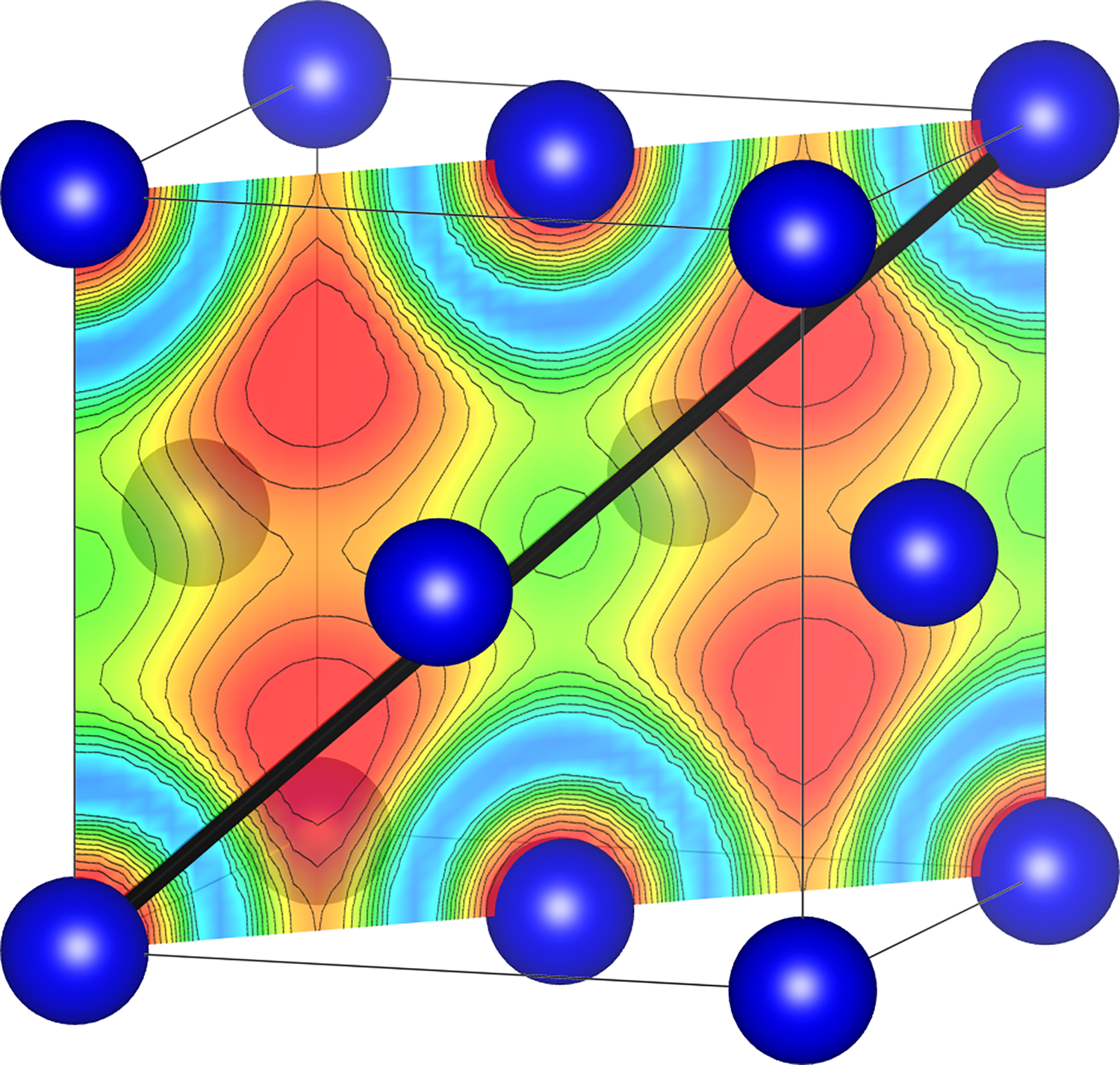}
\includegraphics[width=3.6cm]{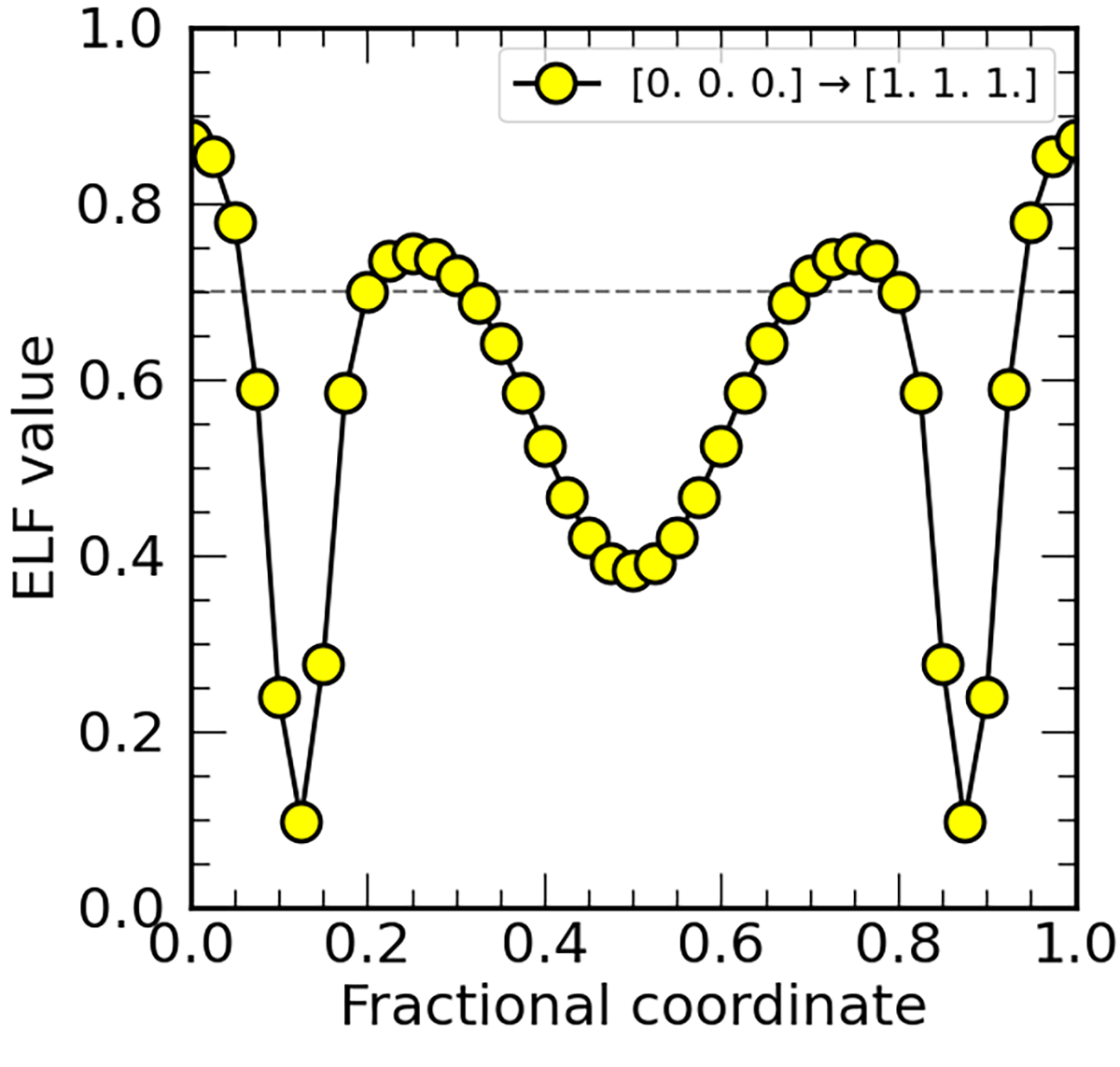}
\caption{Top: ELF isosurface at a value of 0.7 is shown in yellow for the fcc structure (left) at 500 GPa and for the bcc structure (right) of Si at 3500 GPa. Si bcc has 4 yellow dots on the faces of the cell, each counted as a pocket. While the pockets at 0~K seem connected by the green isosurface, that is due to setting the ELF isovalue at ELF = 0.7. Bottom: ELF contour plot along the (110) plane of the fcc unit cell with the [111] direction highlighted as a black line (left) along which the values of ELF are shown on the right. The ELF basins (pockets) correspond to locations where the ELF reaches a local maximum, ${\rm ELF}_{\rm max}>0.7$, but whose value is lower than at atomic sites.}
\label{fig:Si_elf_screenshots}
\end{figure}

We first investigate the geometry and charge of the charge pockets when atoms are at their ideal lattice positions at T = 0 K. Both Si fcc and bcc show to have non-nuclear attractors with ELF values greater than 0.7 (see Fig.~\ref{fig:Si_elf_screenshots}), which aligns with the first point in our proposed criteria. Si fcc had these charge pockets in the tetrahedral sites of the crystal starting at 400 GPa, similar to the electride geometry of fcc carbon~\cite{martinez-canales_thermodynamically_2012}. Below 400 GPa, the ELF values were less than 0.7, and the critical points did not appear in the charge density grid (Table~S5). In the bcc structure, the pockets were on the faces of the cell, with four on each side.

When taking the average ELF local maxima values that are greater than 0.7 and comparing the ELF local maxima positions with the charge density local maxima positions, we integrated the charge density grid w.r.t the ELF  and took the average number of electrons per ELF pocket. We found that Si fcc ($P \geq$ 400 GPa) fits with the electride criteria we proposed, while the Si bcc structure does not because it had less than 1 electron per pocket (Table~\ref{tab:0K_statistics}). Even though the charge density Laplacian was around -2$\times 10^{-2}$ $ e/\mathrm{bohr}^5$ for the Si bcc charge pockets, there was insufficient charge in them, and therefore Si bcc is not an electride. Martinez-Canales \emph{et al.}~\cite{martinez-canales_thermodynamically_2012} found a similar result in carbon, where the fcc structure was an electride, but at a higher pressure, when carbon transitions to the bcc phase, the electride behavior is no longer present. 

\begin{table}[!hbt]
\begin{tabular}{l l l l l} \hline
\multicolumn{1}{p{1.8cm}}{\centering Element/ structure} 
& \multicolumn{1}{p{1.7cm}}{\centering $\langle\rm ELF_{\rm max}\rangle$}
& \multicolumn{1}{p{1.5cm}}{\centering $\langle\rm {\bf elec.}\rangle$/ \\ pocket}
& \multicolumn{1}{p{1.5cm}}{\centering pockets/ \\atom}
& \multicolumn{1}{p{1.5cm}}{\centering $\langle\rm {\bf elec.}\rangle$/ \\ atom}\\ \hline
Si fcc & 0.82(4) & 2.05(4) & 2.00 & 4.09(8) \\ 
Si bcc* & 0.746(2) & 0.706(1) & 6.00 & 4.22(1) \\ 
K fcc* & 0.72(1) & 0.67(2) & 1.00 & 0.67(2) \\ 
K \emph{tI}19 & 0.964(1) & 1.19(3) & 0.62  & 0.73(2)\\
Na \emph{cI}16 & 0.962(5) & 1.47(2) & 0.59 & 0.87(7) \\ 
Na \textit{hP}4 & 0.987(1) & 1.68(3) & 0.50 & 0.84(2) \\ 
Mg bcc* &  0.7018(4) &  0.3271(2) & 6.00 & 1.962(1) \\
Mg fcc & 0.92(2) & 1.66(6) & 1.00 & 1.66(6) \\
Mg sh & 0.88(1) & 0.956(6) & 2.00 & 1.91(1) \\
Mg sc & 0.92(2) & 1.915(5) & 1.00 & 1.915(5) \\\hline
\end{tabular}
\\
* = not an electride 
\caption{The average ELF maxima (of all non-nuclear pockets), average number of electrons per pocket, the number of pockets per atom, and the total number of pocket electrons per atom. These averages are taken from all simulated structures within the pressure ranges from Table~\ref{tab:MD_table} at $T = 0$~K. 
The data presented for Si in the fcc structure corresponds to simulations at $P \geq$ 400 GPa, where electride behavior begins to emerge. 
\revisiontwo{We did not include K bcc in this table because all ELF maxima were below 0.7.}
}
\label{tab:0K_statistics}
\end{table}

\subsubsection{Sodium and Potassium}
\begin{figure*}[!hbt]
    \centering
    \includegraphics[height=3.4cm]{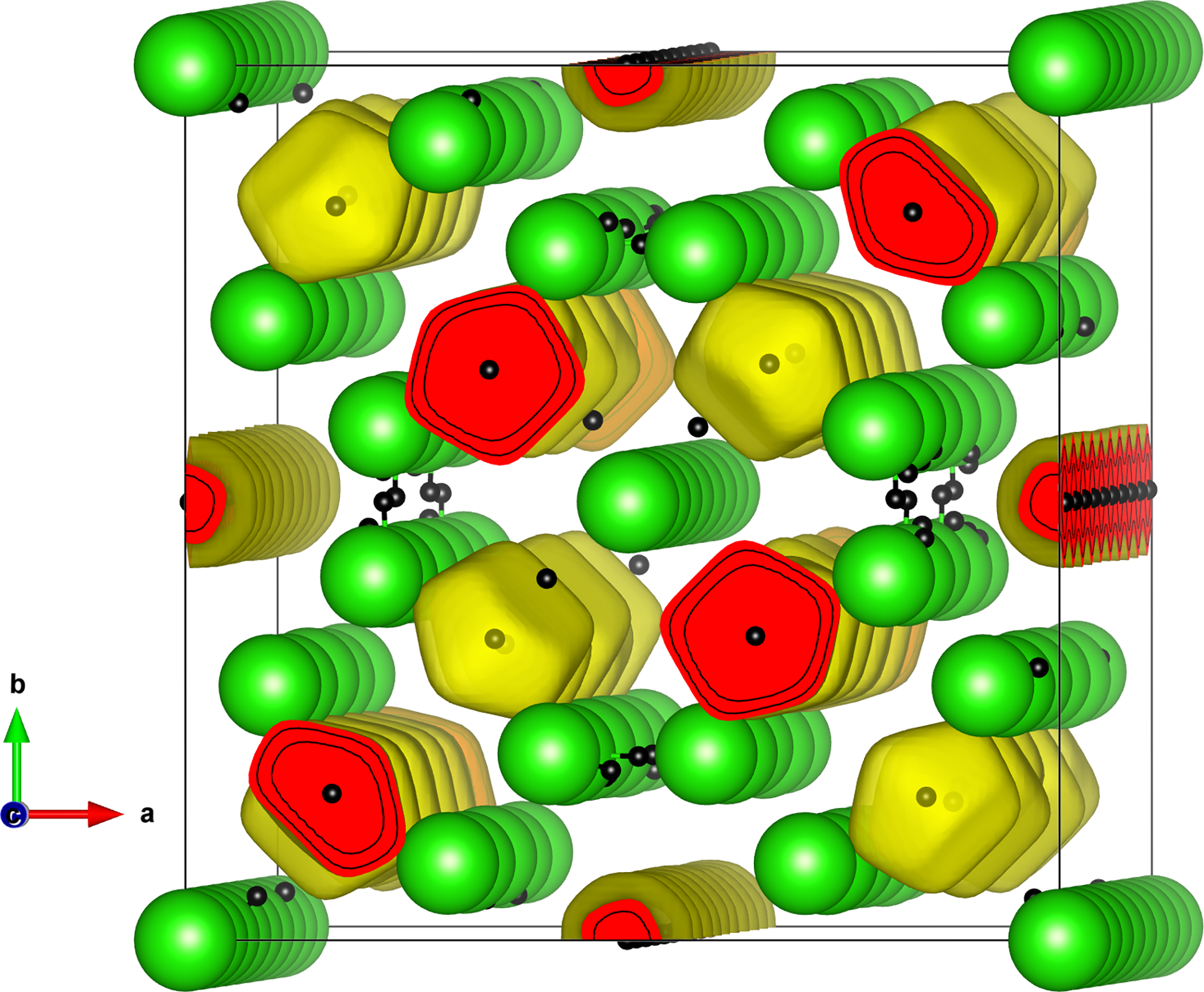}
    \includegraphics[height=3.4cm]{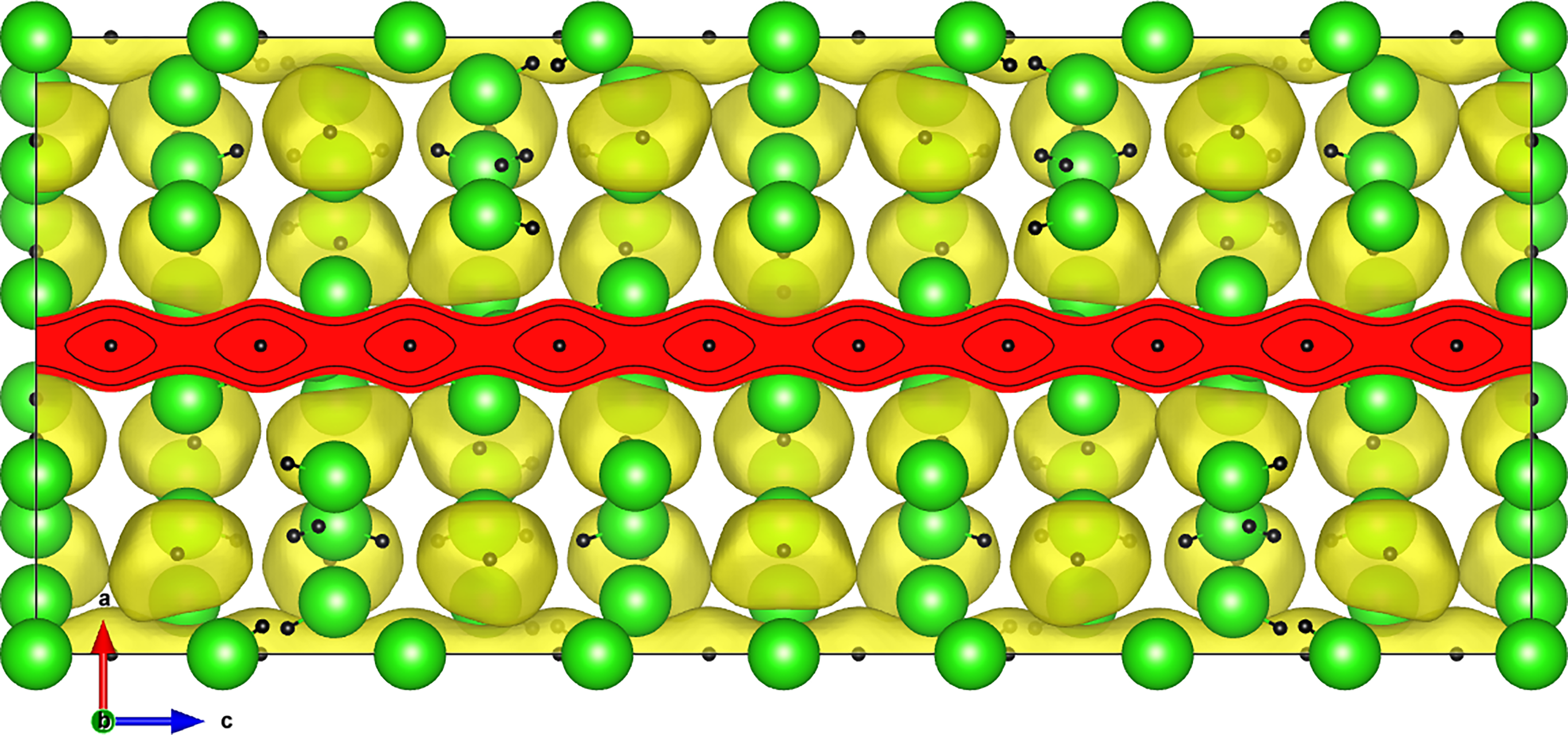}
    \includegraphics[height=3.4cm]{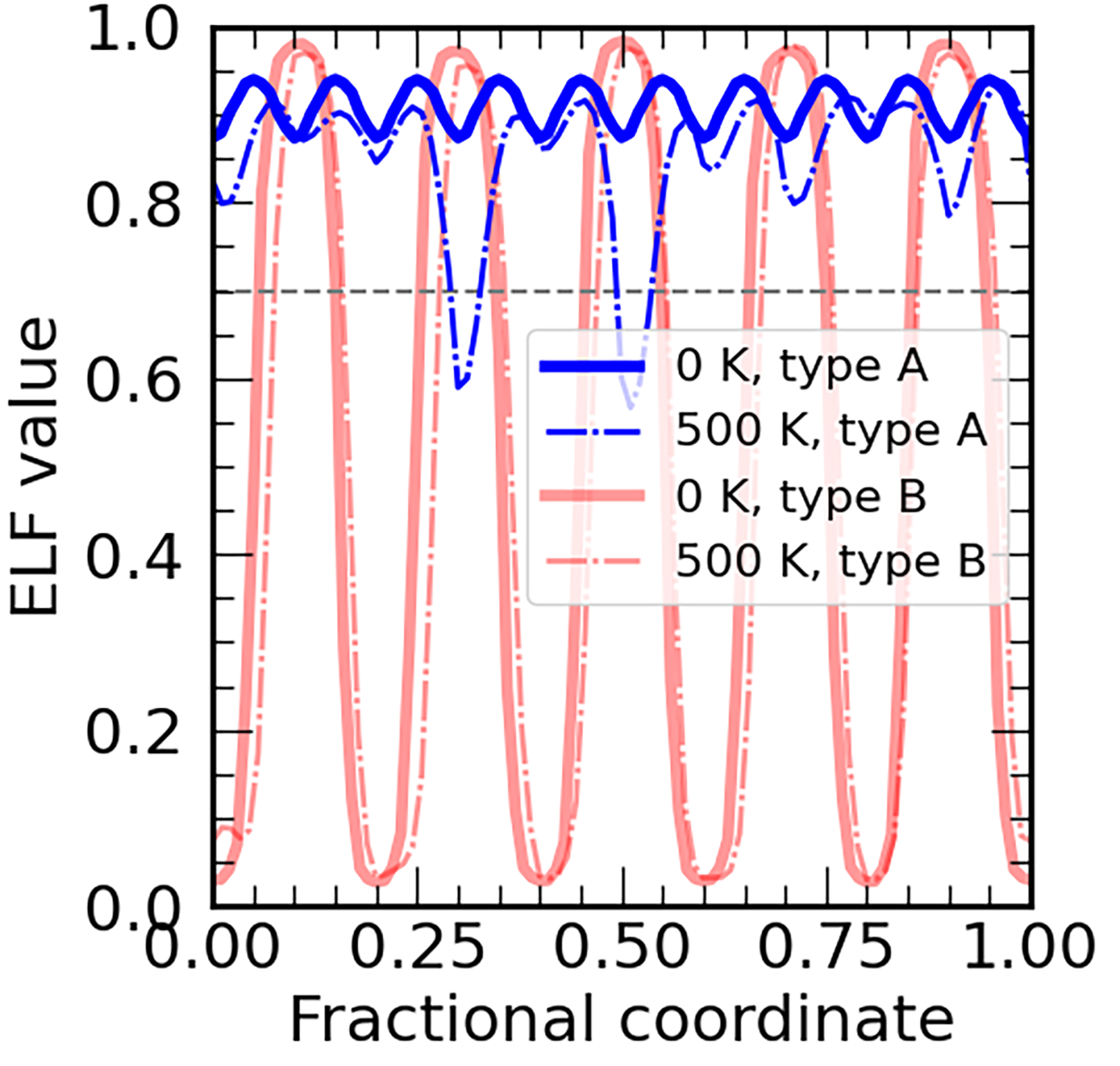}
    \caption{Left and middle: Unit cell of the \emph{cI}16 structure of sodium at 200 GPa with Na atoms colored in green, critical points at non-nuclear sites in black, and ELF isosurface in yellow. Right: a line profile along the [001] crystalline direction starting at the point (0, 1/2, 0) in blue (along type A maxima), and starting from the point (0.34288,0.34288,0) in red (along type B maxima). At 500 K, the ELF values at both type A and B maxima barely change, but they can decrease strongly in between type B maxima, indicating more delocalization between these pockets.}
    \label{fig:Na_line_profile_main}
\end{figure*}

Na and K both exhibit electride behavior in the incommensurate host-guest structure, which contains two distinct types of charge-localization regions~\cite{woolman_structural_2018}: more spatially extended ELF pockets along the host channels (which we will call ``channel A") and more confined ELF pockets along the guest channels (which we will call ``channel B"), which lie inside the host structures. Both channels form 1D atomic guest chains located along the (110) and ($\bar 1 10$) planes (see Fig.~\ref{fig:Na_line_profile_main}). We simulated Na \emph{cI}16 at 160 GPa in this host-guest structure, and when we integrated the charge contained in the ELF basins that lie along channel B, we found that the average charge per ELF pocket was 0.98 electrons. The basins along channel A, which are larger by volume, contain 1.61 electrons each. Thus, on average, there are 1.5 electrons per ELF pocket, but this value does not take into account that they come in two different populations. These values are in agreement with those reported by Woolman et al., who found 1.69 electrons in the channel A ELF maxima and 1.04 electrons in the channel B ELF maxima. When we evaluate the ELF function along the path that connects the channel A maxima (through the center of the host structure), which is the [001] crystalline direction starting at the point (0, 1/2, 0), we obtain the line profile shown in blue on the right of  Fig.~\ref{fig:Na_line_profile_main}. As we can see, a temperature of 500 K can significantly affect the ELF values found along this path and, consequently, the shape of basins where electrons localize. To the contrary, the channel B maxima are significantly more stable with respect to temperature, as they remain with high values and well separated even when atoms are vibrating.

Na hP4 was one of the first high-pressure electrides studied, discovered by Ma \emph{et al.}~\cite{ma_transparent_2009}, and it has since been the model structure of an electride due to its high electron localization. Racioppi \emph{et al.}~\cite{racioppi_electride_2023} quantified the number of electrons in each charge pocket using the charge density as the limits of integration rather than the ELF. They found 1.07 electrons in each basin, and while we find 1.7 electrons w.r.t to the ELF basin topology, we find that Na hP4 had 1.11 electrons when integrating in reference to the charge density itself, which agrees with previous literature (Table~S8).

Before reaching its electride state, K was simulated in the bcc and fcc structures. K bcc did not have any ELF values above 0.7, meaning that all electrons are delocalized. This is expected considering that K bcc is metallic at ambient pressures. K fcc, however, did have ELF maxima greater than 0.7 starting at 13 GPa, but the charge in the maxima integrated to only 0.67 electrons and therefore is still not an electride. In addition, K fcc had a charge density Laplacian in the order of $10^{-4}$ $ e/\mathrm{bohr}^5$, which is also below our proposed threshold of $10^{-3}$ $ e/\mathrm{bohr}^5$. This means that the charge pockets we see in K fcc are more likely to be bonds rather than charge pockets associated with electrides. 

Both Si bcc and K fcc are non-electrides that exist at pressures either before or after the electride phases. As such, we see ELF values above 0.7, highlighting the localized charge that exists in non-electrides at high pressures. At the same time, the ELF alone cannot classify electride behavior, for the highly localized behavior can come from other sources such as covalent bonding. In addition, the local maxima in the ELF basins should have sufficient charge (i.e., at least 0.9 electrons) to ensure strong localization. Otherwise, any non-nuclear charge can be classified as an electride.


\subsubsection{Magnesium}
\begin{figure}[!hbt]
    \centering
    \includegraphics[width=3.7cm]{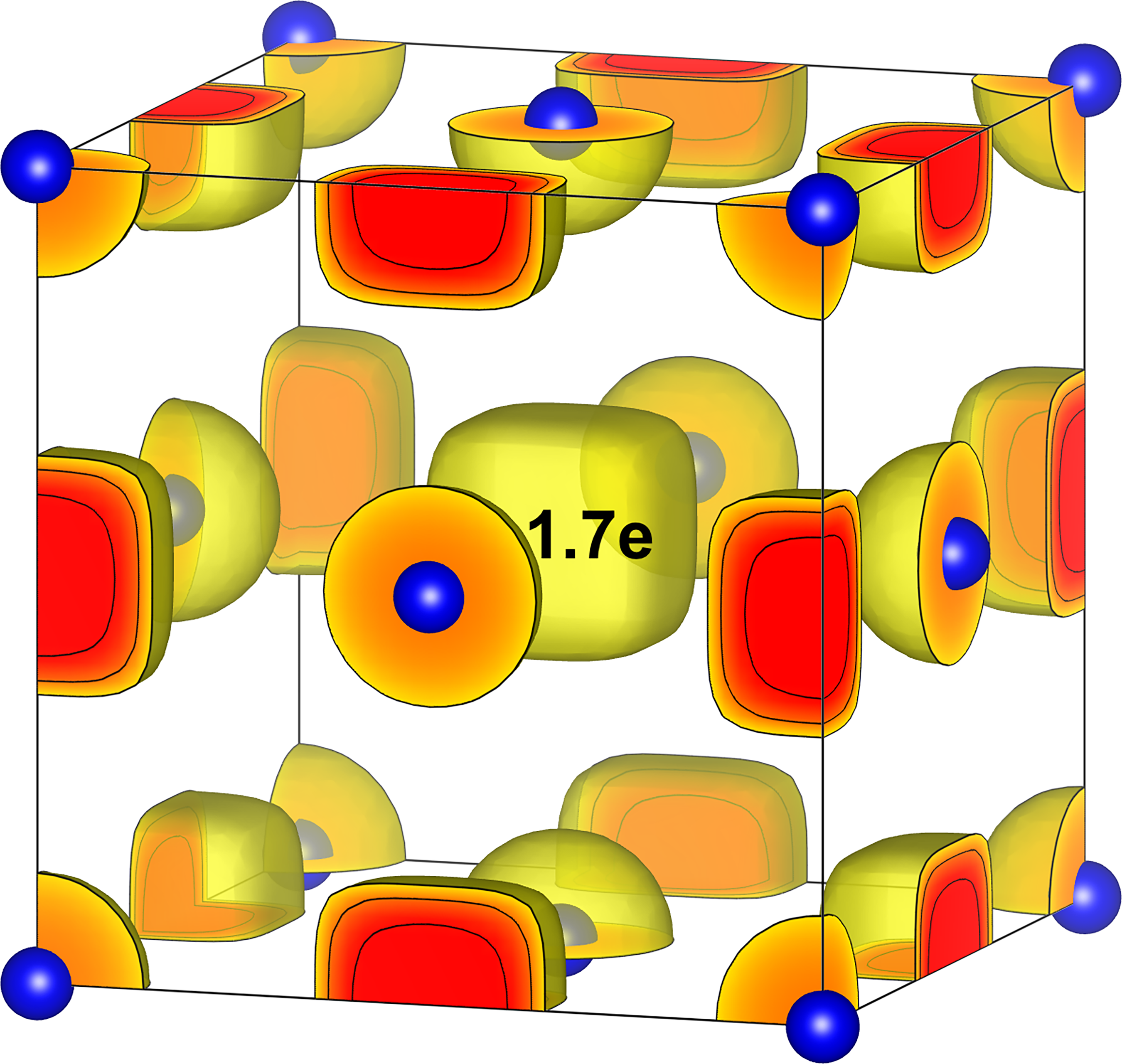}
    \includegraphics[width=3.7cm]{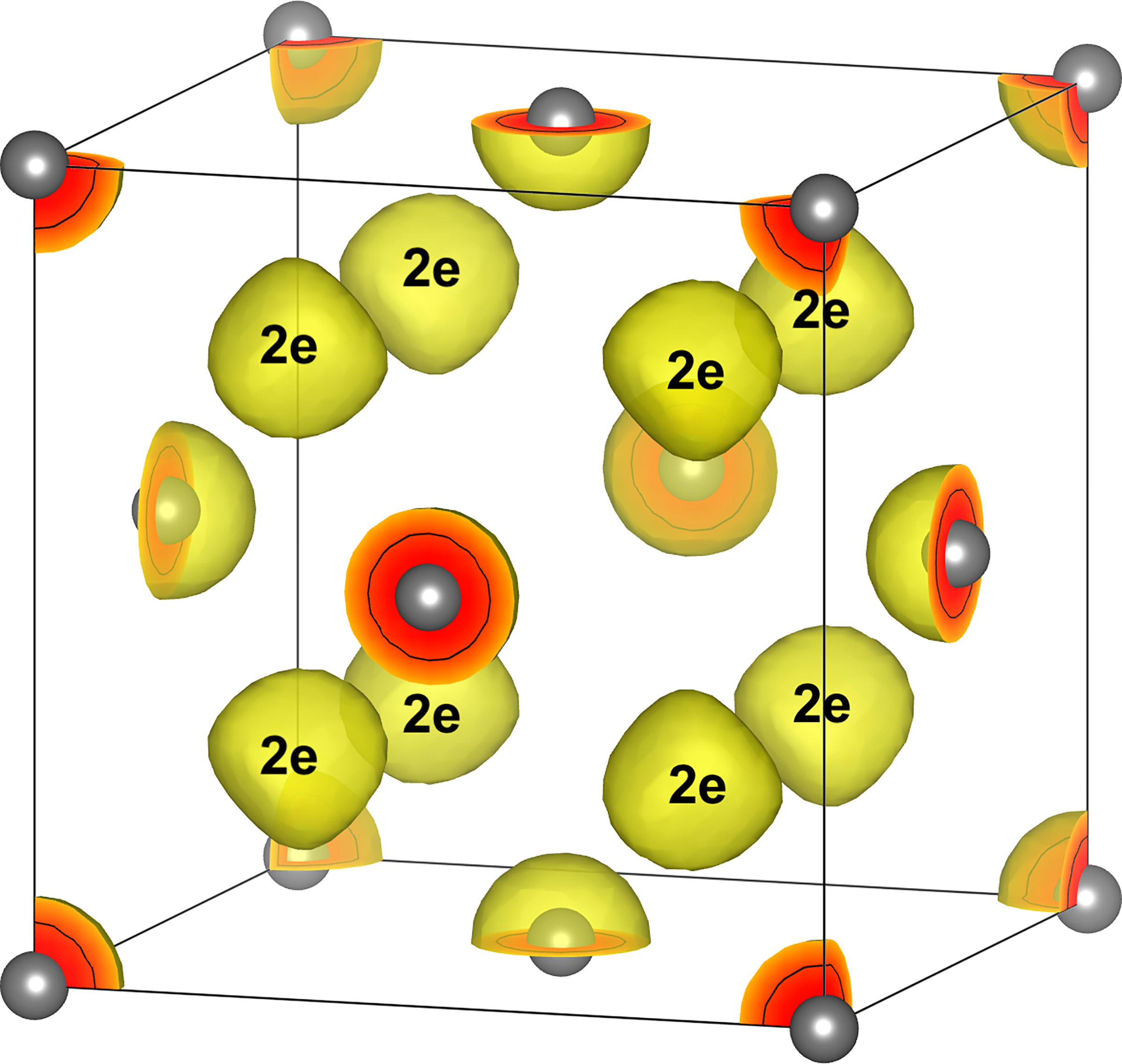}
    \caption{Electronic localization function (ELF) in Mg (left) and Si (right) in the fcc structure. 
    In Mg (1000 GPa), charge basins contain a charge of 1.7 electrons and are located at octahedral positions along the cube edges, while in Si (500 GPa), the charge basins are inside the cube in the tetrahedral sites and contain two electrons.}
    \label{fig:FCC_Mg_vs_Si}
\end{figure}

Electride behavior in magnesium was first discovered computationally by Li \emph{et al.}~\cite{Li_Mg_electride_2010} for Mg fcc and Mg sh, and later Mg sc by Gonz\'alez-Cataldo~\emph{et al}~\cite{gonzalez-cataldo_structural_2023}. Experimental observations of high-pressure phases of elemental magnesium have confirmed these phase transitions~\cite{gorman_experimental_2022}. Using the ramp compression technique~\cite{Gonzalez-Cataldo2021}, Gorman \emph{et al.} demonstrated that magnesium transitions from hcp to bcc and then to fcc at pressures above 450 GPa. At even higher pressures, the simple-hexagonal (sh) and simple-cubic (sc) phases emerge. In the fcc phase, electron localization occurs at octahedral positions relative to the face atoms of the cell. In Fig.~\ref{fig:FCC_Mg_vs_Si}, we compare this localization in Mg with Si in the same fcc phases at the ELF isosurface value of 0.7.
We can also observe that the electron localization pattern of Mg is topologically distinct from that of Si in the same fcc phase, which is stable around the same pressure range but shows electron localization at tetrahedral sites. For Mg, there are four localized electron sites per unit cell, while Si can accommodate eight of these pockets in the same structure. Given that the fcc unit cell contains four atoms, this distribution results in one localized electron pocket per atom for fcc-Mg and two pockets per atom for fcc-Si. 
However, an analysis of the electron charge contained within these pockets shows that each of them contains approximately 1.7 electrons in Mg, while each pocket in Si contains 2 electrons, as shown in Fig.~\ref{fig:FCC_Mg_vs_Si}. This aligns with their valence configurations: each Si atom, with four valence electrons ([Ne]3s$^2$3p$^2$), contributes with these four electrons fully localized into two pockets, whereas Mg, with two valence electrons ([Ne]3s$^2$), places roughly 1.7 electrons per pocket.

\begin{figure}[!hbt]
    \centering
    \includegraphics[width=0.9\linewidth]{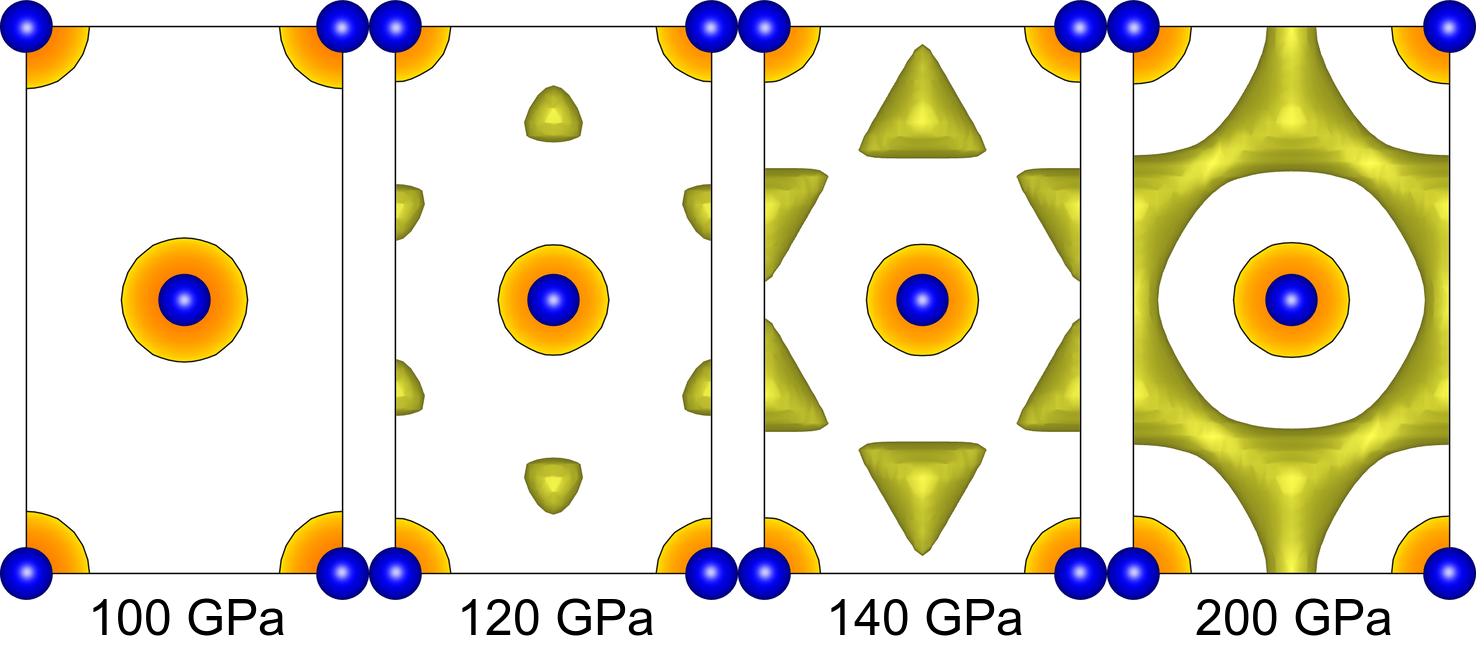} \\
    \includegraphics[width=0.9\linewidth]{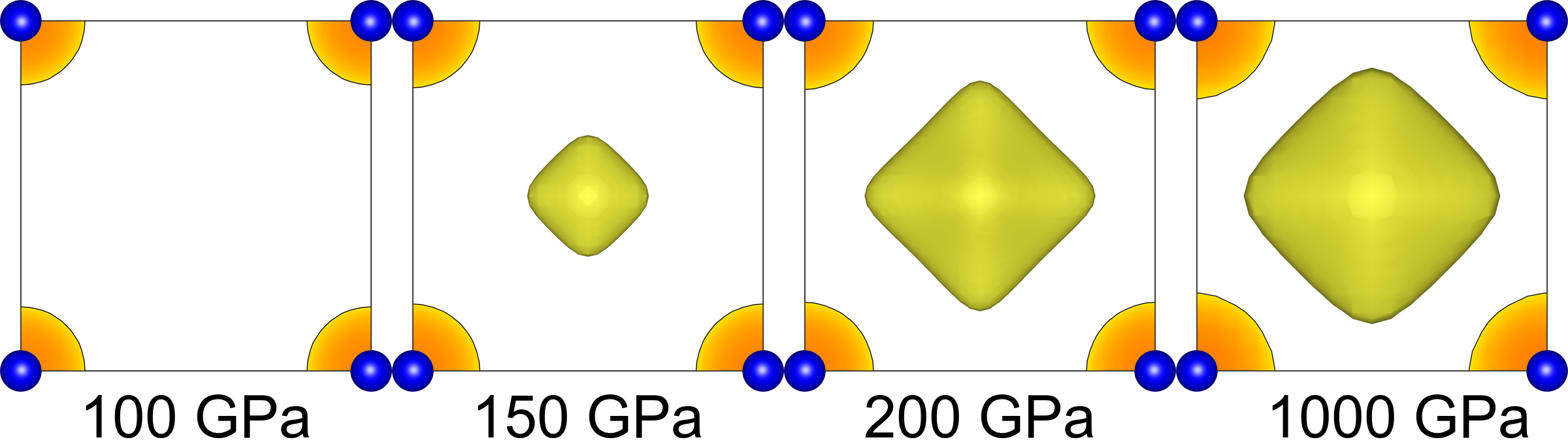} \\
    \caption{
    Pressure-induced localization of electrons in simple-hexagonal (top) and simple-cubic (bottom) Mg. All panels depict the isosurface defined by the ELF value of 0.7. 
    }
    \label{fig:Mg_SH_and_SC_evolution}
\end{figure}

A similar behavior is observed for the higher-pressure phases of Mg, the simple-hexagonal (sh) and simple-cubic (sc) phases, depicted in Fig.~\ref{fig:Mg_SH_and_SC_evolution}. As we can observe in the figure, pockets of localized electrons start forming well below the pressure where each structure becomes thermodynamically stable~\cite{gorman_experimental_2022}. As pressure increases, the charge pockets in the sh structure appear to connect to create a ring-shaped, connected structure. The sh structure, however, still has 6 isolated pockets if a higher ELF value is used to display the isosurface. This ``merge" in Mg sh differs from the heated Si bcc solid in Fig. \ref{fig:Si_stats_vs_pressure}. Rather than the number of critical points changing like in Si bcc, we see the same number of critical points for Mg sh as pressure increases, except the points have more ELF values greater than 0.7. In other words, this continuous isosurface we see for Mg sh at pressures $\geq$ 200 GPa is simply an artifact of ELF visualization. This further highlights the importance of studying the gradient of the ELF and charge density grids. 
We propose that a critical point should have an ELF peak maximum greater than 0.7 to be a charge pocket. We use the gradient and surrounding zero-flux regions in the ELF grid to define the limits of integration. Finally, we integrate the charge density grid using the limits of integration set by the ELF grid to quantify the amount of charge in each charge pocket. 
Since there are 4 charge pockets in the sh unit cell and 2 atoms per unit cell, we have 2 charge pockets per atom. When we analyze the charge density in these pockets that appear in the sh structure, we find that they contain approximately 1 electron each.

For the simple cubic structure in Fig.~\ref{fig:Mg_SH_and_SC_evolution}, only one pocket is developed with pressure, which contains almost 2 electrons (see Table~\ref{tab:0K_statistics}). Thus, the ELF basins in these three high-pressure structures of Mg (fcc, sh, and sc) can accommodate between 1 and 2 electrons in each pocket, but the total localized charge per atom is 2e in all Mg structures, except for the bcc structure, which is not an electride. Interestingly, the fcc and sc phases of Mg exhibit 1 charge pocket per atom, while in the sh there are 2. Nevertheless, regardless of the number of pockets formed, there are around 2 electrons per atom that are contained in pockets.

In summary, Si fcc is an electride at pressures greater than 400~GPa, and Si bcc is not an electride because it has only 0.71 electrons per pocket. The other electride structures agreed with our quantitative electride criteria in
while K bcc, K fcc, and Mg bcc did not because they are not electrides. When comparing structures for each element, we found that the same amount of charge (i.e., electrons per atom) is consistent, and more so aligns with the number of valence electrons each element has. For example, Si has 4 valence electrons, and we found that both Si fcc and bcc had 4 electrons per atom that were in the interstitial sites. Even though K fcc is not an electride and K \emph{tI}19 is, both had a total of around 0.7 electrons in all charge pockets per atom, which is close to the 1 valence electron that K has. Furthermore, the amount of charge that is in the interstitial sites tends to remain constant, but the distribution of the charge is what changes and therefore dictates whether a crystal is an electride or not.

\subsection{Solid and liquid electrides at high temperature}
In this section, we discuss how temperature affects the electride behavior of each material and compare it with their respective electron localization at 0~K. We consider solids heated below the melting temperature and liquids above the melting temperature of each material.

\subsubsection{Silicon}
\begin{figure*}[!hbt]
    \centering
    \begin{subfigure}[b]{0.62\linewidth}
        \includegraphics[width=10cm]{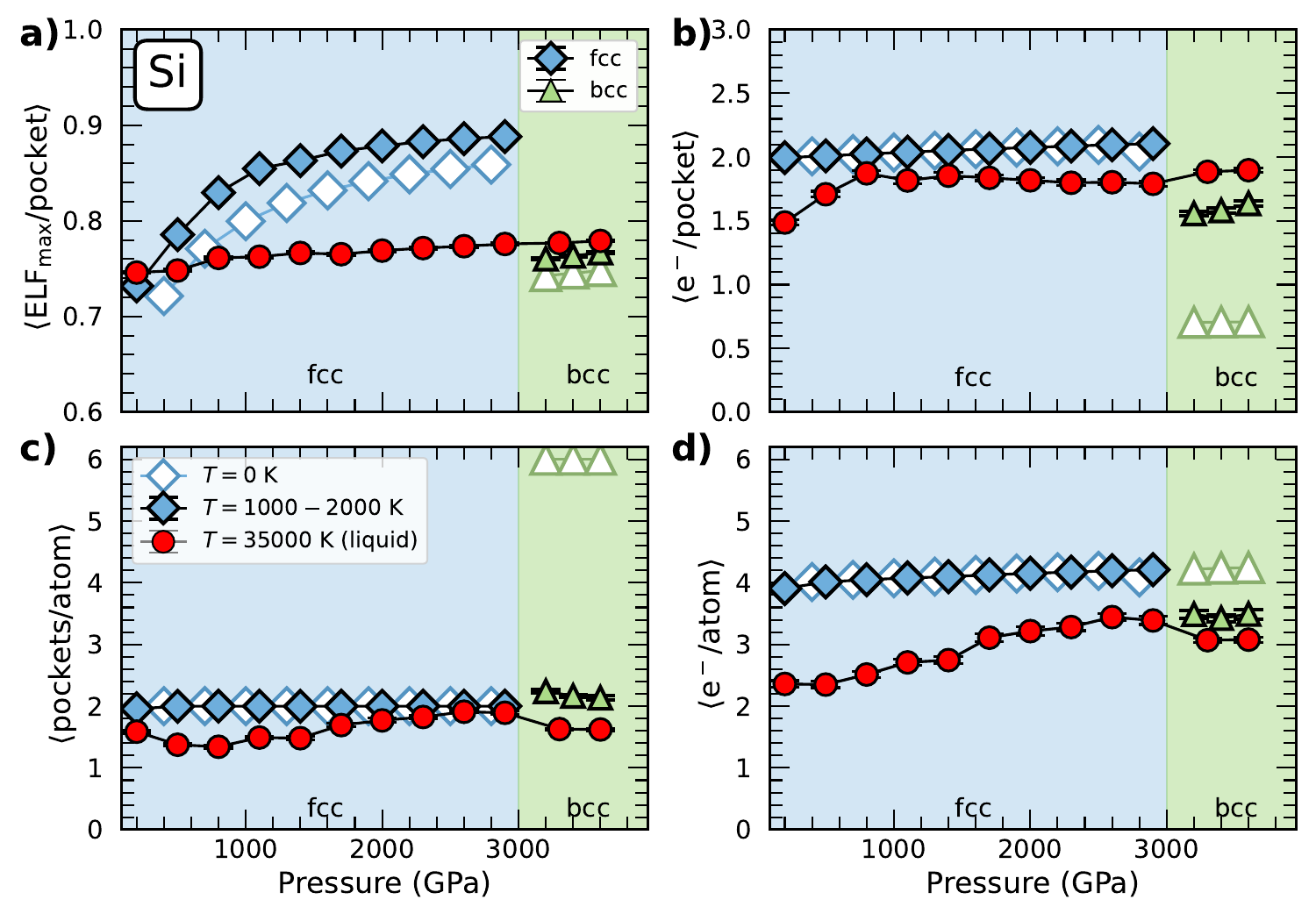} 
    \end{subfigure}
    \begin{subfigure}[b]{0.35\linewidth}
        \centering
        \includegraphics[width=2.8cm]{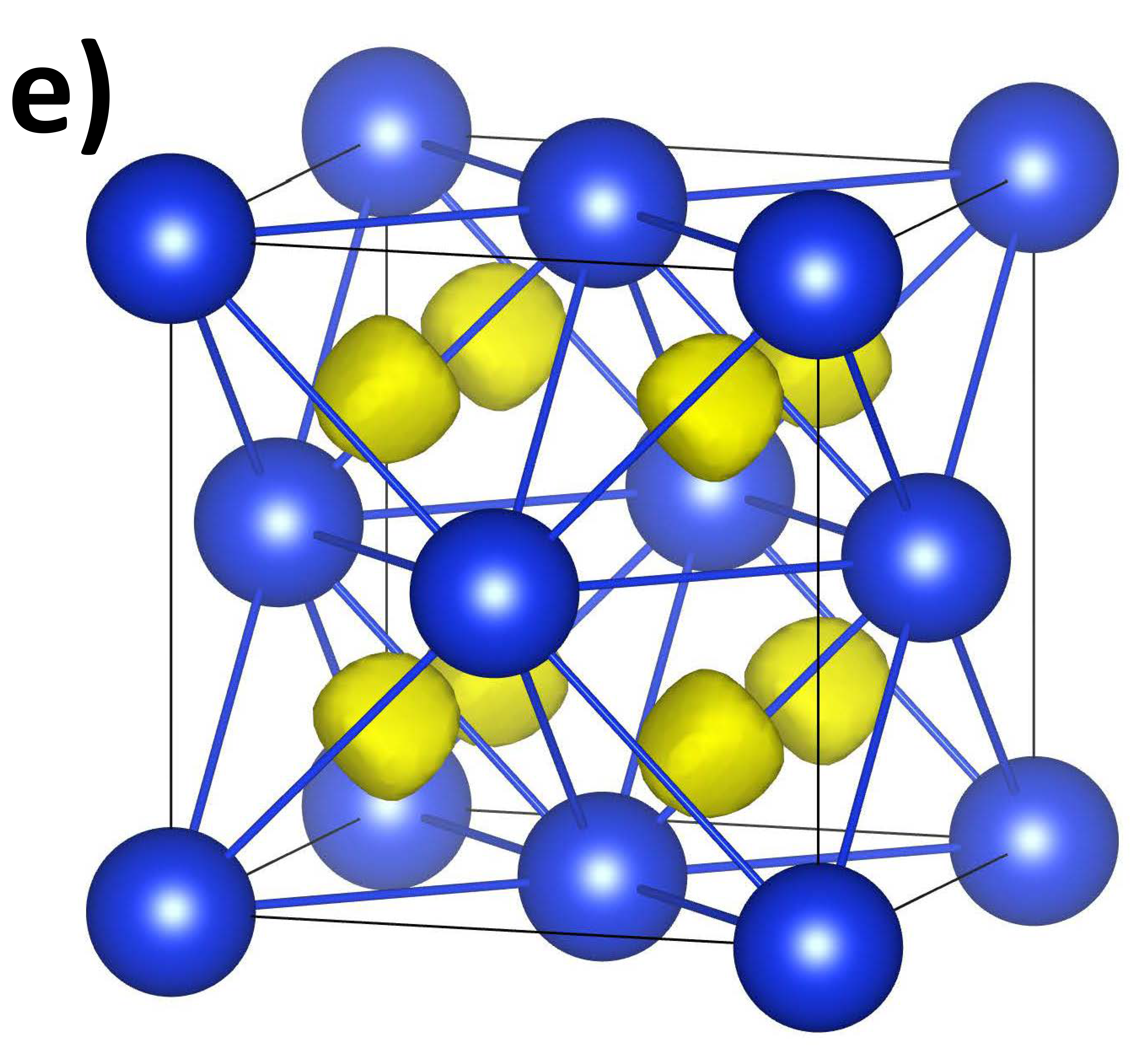}
        \includegraphics[width=2.8cm]{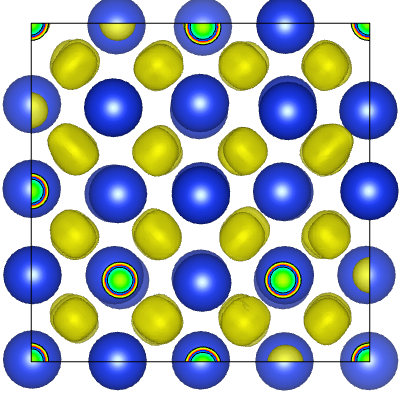}\\~\\
        \includegraphics[width=5.4cm]{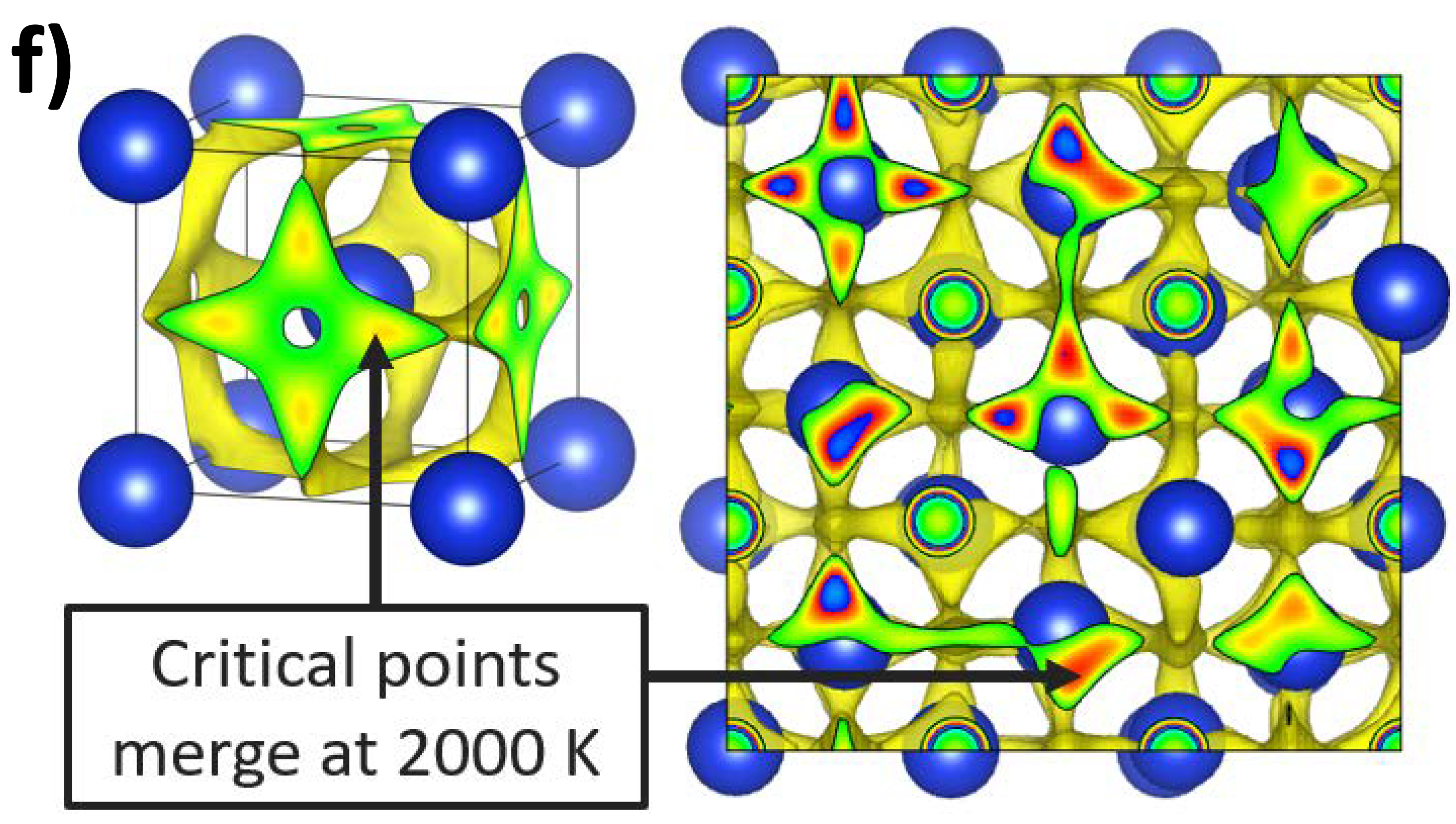} \\~\\
    \end{subfigure}
    \caption{Silicon: 
    (a) The average ELF maximum value of each pocket, (b) the average number of electrons per pocket, (c) the average number of pockets per atom, and (d) the average number of electrons in charge pockets per atom vs pressure for Si (starting at 200 GPa). The different shading represents the different stable structures in those pressure regimes. The transparent points represent the T = 0K data, the filled-in shapes matching their backgrounds represent the heated solids, and the red circles represent the melted structures (i.e. liquids). The standard error bars for most data points are smaller than the markers.
    e-f) 3D ELF (isovalue $=$ 0.7) of Si (blue atoms) from top to bottom of fcc and bcc, respectively. e) Si fcc unit cell at 0 K (left), and the heated super cell (right). f) The Si bcc unit cell at 0~K and heated super cell at 2000~K. 
    There were 4 uniform pockets on each face of the cell at 0~K, represented by the 4 yellow dots. This equates to 6 pockets per atom for Si bcc at 0~K. While the pockets at 0~K seem connected by the green isosurface, that is due to setting the ELF isovalue at ELF = 0.7. At 2000~K, some of those pockets merge, going from 6 pockets per atom to around 2 (plot c). The ELF pockets on the faces of the bcc cells have colors representing the relative ELF intensities, increasing from blue to green to red.}
    \label{fig:Si_stats_vs_pressure}
\end{figure*}

We heated Si fcc phase at 1000 K and Si bcc at 2000 K (Table~\ref{tab:MD_table}) to investigate how electride behavior in solids changes when the atoms move.
In Fig.~\ref{fig:Si_stats_vs_pressure}, we show the Si fcc and bcc unit cells together with their respective supercells heated to 1000~K and 2000~K, respectively, with the charge pockets (yellow) defined by an ELF value of 0.7.
As we can observe in the figure, the solids at high temperature exhibit ELF pockets slightly deformed by the atomic vibrations, but still visible. 
As pressure increases, the ELF value at the center of the pockets depicted in Fig.~\ref{fig:Si_stats_vs_pressure} increases, indicating stronger electron localization. 
For Si fcc, we found that while the ELF increased, the number of electrons per pocket and per atom did not change. As temperature increased from 0~K to 1000~K for a given pressure, there is a jump in the ELF local maximum value per pocket, and then it decreases upon melting. Because of this jump in ELF from 0~K to 1000~K, we see Si fcc having ELF values greater than 0.7 as low as 200 GPa. 
The amount of charge per pocket and the number of pockets per atom did not change when Si fcc was heated at 1000~K, but decreased upon melting (Table S4, Fig. \ref{fig:Si_stats_vs_pressure}). 
This implies that electride behavior has a stronger correlation with structure rather than pressure alone.
Yes, pressure does dictate structure stability and orbital overlap, but we find a greater change in charge per pocket when changing the structure while keeping pressure constant (i.e., melting the structure) rather than vice versa.

Unlike Si fcc, the charge pocket geometry for Si bcc was more sensitive to temperature. The number of electrons per pocket increased by a factor of 2 once Si bcc was heated, but the number of pockets per atom was reduced by a factor of 3. This implies that the charge pockets found at T = 0 K merged once the atoms moved slightly out of their ideal positions (while maintaining the bcc structure). Consequently, we saw a less uniform distribution of the charge pockets for Si bcc when heated (Fig. \ref{fig:Si_stats_vs_pressure}), compared to Si fcc. 
Upon melting, we see this trend continue, but only slightly. Si, as a liquid, had less charge per pocket than Si fcc when heated, but had more charge per pocket than Si bcc when heated. Si liquid still satisfies the electride criteria for it had ELF maxima $\geq$ 0.7, more than 0.9 electrons per pocket, and charge density Laplacian values in the order of $10^{-2}$ or larger.
Overall, Si fcc maintains its electride behavior when heated, but decreases upon melting. In contrast, Si bcc is initially not an electride at T = 0 K; it becomes an electride when thermal disorder occurs, and then the electride behavior increases slightly upon melting. Electride behavior for Si as a liquid stays relatively the same across all pressures.

\subsubsection{Sodium and Potassium}
Na was heated at 300 K for the \emph{cI}16 structure and to 500 K for the hP4 structure. Both of these structures are electrides that have been previously studied~\cite{ma_transparent_2009,woolman_structural_2018}. We found that all the solid structures agreed with our quantitative thresholds. Similar to Si, both solids had the same amount of charge in the ELF pockets when heated, but decreased when melted (Fig.~S6). 
Like Si fcc, Na hP4 has its charge pockets located in the tetrahedral sites of the lattice. Even though both Na structures are known electrides, both had less than 1 electride electron \textit{per atom}. 
Na hP4, for example, only has 2 ELF pockets per unit cell, which has 4 atoms. As a result, there will be 0.5 pockets per atom (Table \ref{tab:0K_statistics}). So if there are around 1.6 atoms per pocket, that would result in 0.8 electrons per atom. Nevertheless, both Na hP4 and \emph{cI}16 had more than 0.9 electrons \textit{per pocket}, and therefore both satisfy our proposed electride criteria.

\begin{figure}[!hbt]
    \centering
    \includegraphics[width=2.5in]{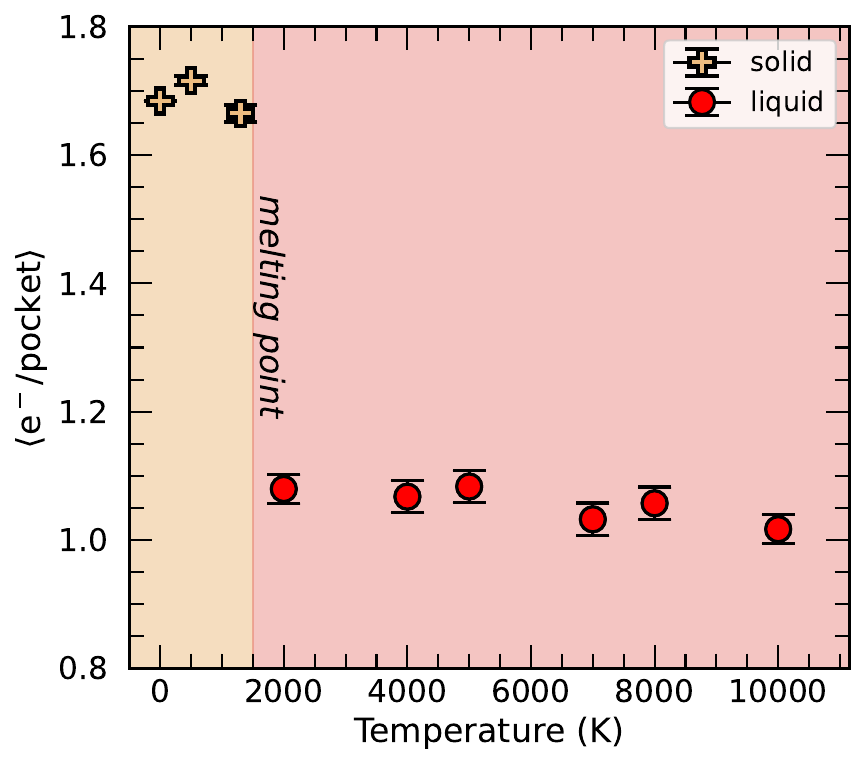} \\
    \includegraphics[width=2.5in]{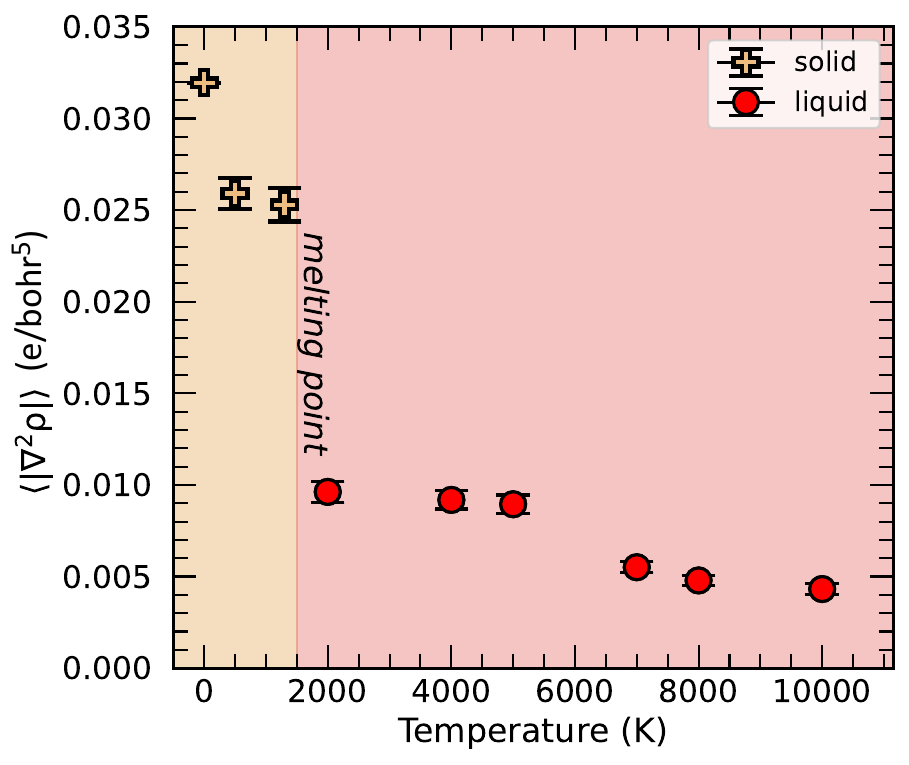}
    \caption{Sodium: (top) Number of electrons per pocket and (bottom) the average absolute value of the negative charge density Laplacian of the pockets as a function of temperature for sodium in the hP4 and liquid phase at $\sim$250 GPa (at constant density = 5.367 g/$\rm{cm}^3$). The approximated melting point is at $T = 1500$~K. 
    }
    \label{fig:Temp_dependence_plot}
\end{figure}

To further investigate the relationship between electride behavior and temperature, we heated Na hP4 from 0 K to 10,000 K and plotted the charge per pocket at each temperature (Fig. \ref{fig:Temp_dependence_plot}). There was a sudden drop from 1.8 to approximately 1 electron per pocket at the melting line. We also see a drop with Laplacian values going from $|\nabla^2 \rho(\mathbf{r}_0)| \simeq 10^{-2}$ to $|\nabla^2 \rho(\mathbf{r}_0)| \simeq 10^{-3}\ e/\mathrm{bohr}^5$. 
Paul \emph{et al.}~\cite{paul_thermal_2020} also investigated the relationship between electride behavior and temperature for Na hP4. Upon melting, they saw the disappearance of the band gap that Na hP4. While Na is an electride in both the solid and liquid phases, the loss of the band gap led to the charge pockets becoming smaller with less charge. Therefore, we find that Na becomes a weaker electride when melted because there is less charge in the interstitial sites and because it's no longer fully insulating (i.e., there's no longer a band gap). 

\begin{figure}[!hbt]
\centering\hskip-10mm
\includegraphics[width=8cm]{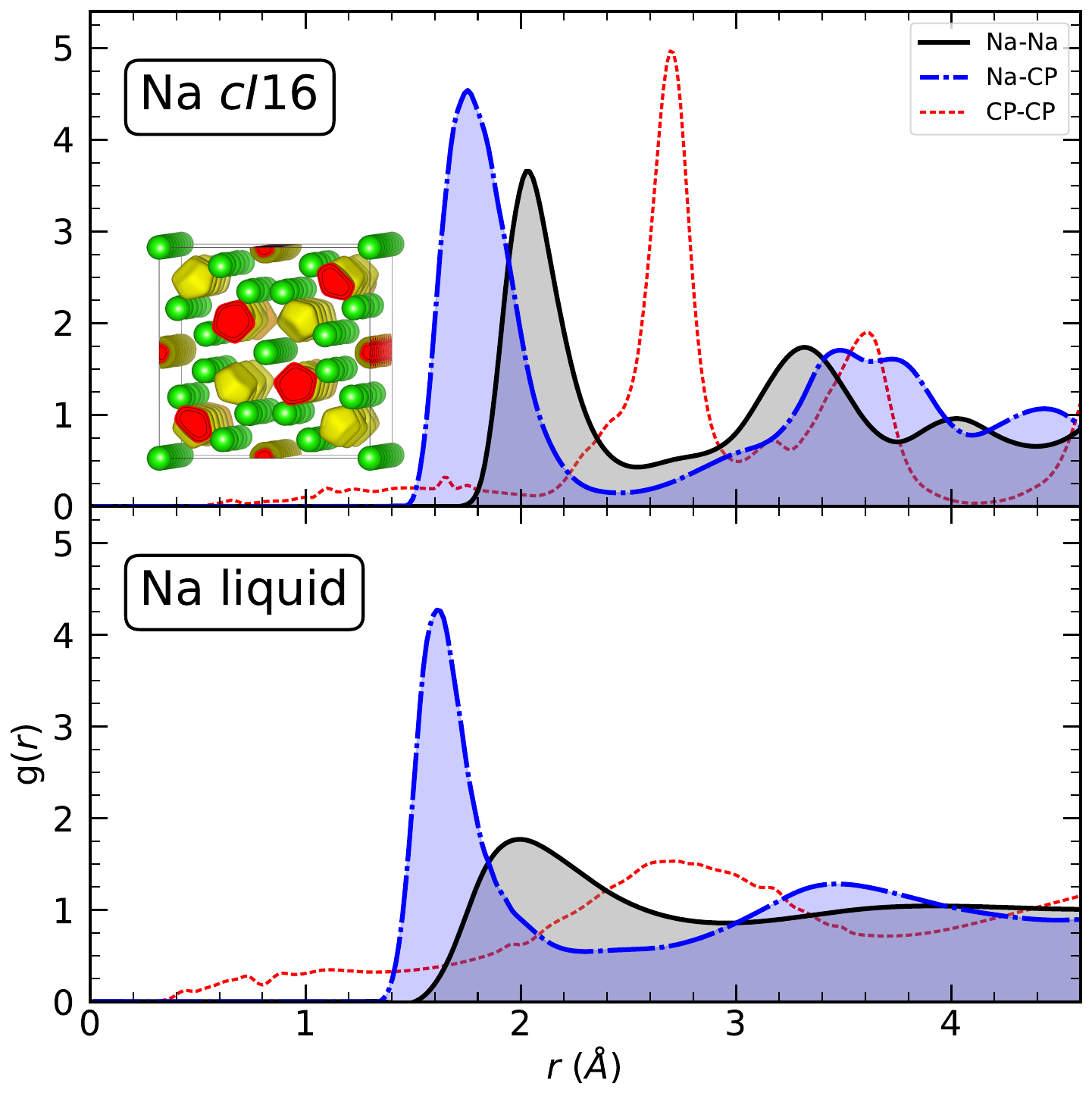}
\caption{Nuclear (Na) and critical points (CP) pair correlation functions computed with DFT-MD simulations of sodium at $\sim$200~GPa (4.8 g cm$^{-3}$). Top: Solid sodium in the \emph{cI}16 phase at 500 K. Bottom: Liquid sodium at 3000~K and same density. The Na-CP correlation shows that Na atoms are closer to electron pockets than to other Na atoms.}
\label{fig:gor_Na}
\end{figure}

The positions of the non-nuclear critical points (CP) associated with localized electron pockets with ${\rm ELF}>0.7$ can be treated as additional particles, and the pair correlation functions can be computed. Fig.~\ref{fig:gor_Na} shows these correlations for Na in the solid (\emph{cI}16) and liquid phases at $\sim$200~GPa. In solid Na, the location of the first Na-Na peak does not coincide with the first CP-CP peak, indicating that the nearest-neighbor distance between Na atoms is smaller than the distance between two nearby electron pockets, as shown in the unit cell inset. This suggests that the solid at high temperature (500~K) retains the same spatial distribution of electron pockets as at 0 K, remaining stable despite thermal vibrations. The first Na-CP peak is located at even smaller distances, indicating that Na atoms are closer to electron pockets than they are to other Na atoms. The structure of the liquid retains a very close resemblance to the solid, with both electron pockets and Na atoms separated by basically the same average distances from each other as they are in the solid. It is interesting to note that the second and third Na-CP peaks of the solid have merged into a single secondary peak in the liquid phase, meaning that the liquid has lost some of the long-range correlations between atoms and electron pockets but still preserves the short-range structural organization of the solid.

\begin{figure}[!hbtp]
    \centering
    \hspace{0.8cm}
    \includegraphics[width=1.8cm]{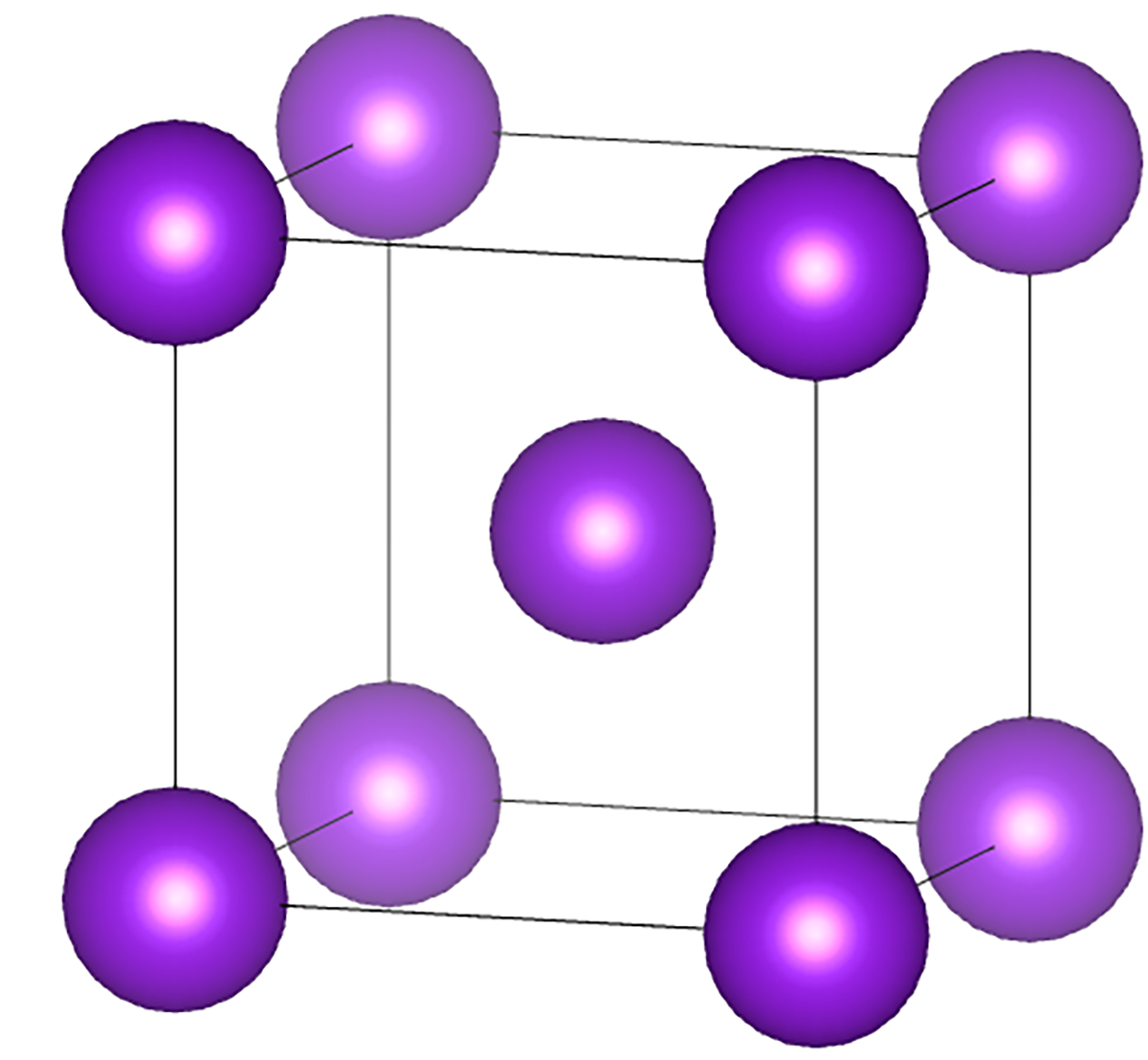}
    \includegraphics[width=1.8cm]{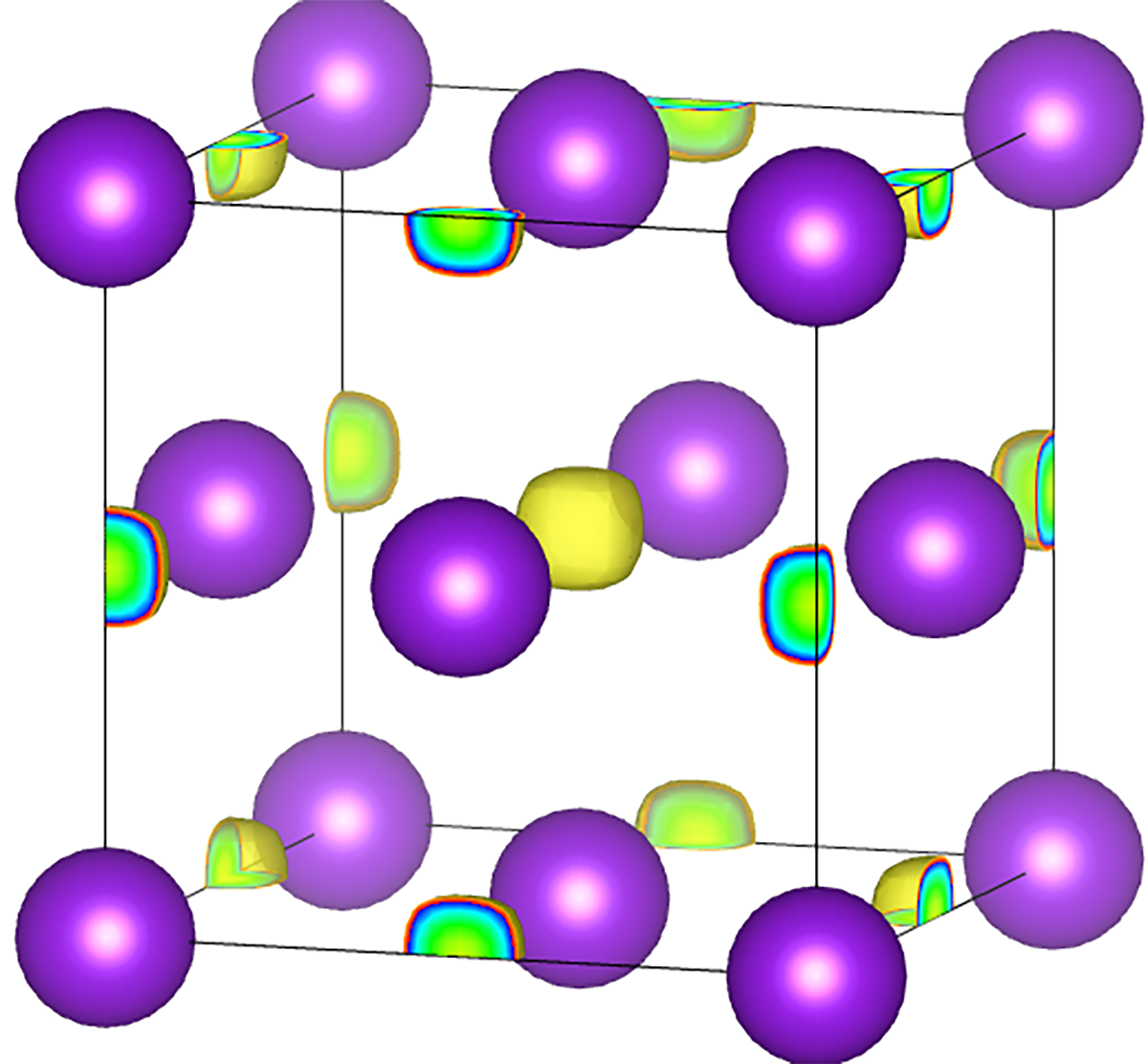}
    \hspace{0.1cm}
    \includegraphics[width=1.7cm]{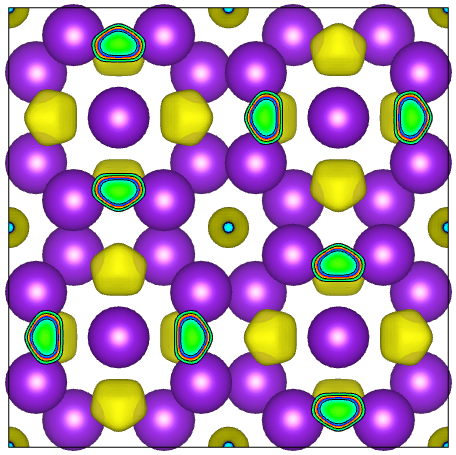} \\
    \includegraphics[width=6.6cm]{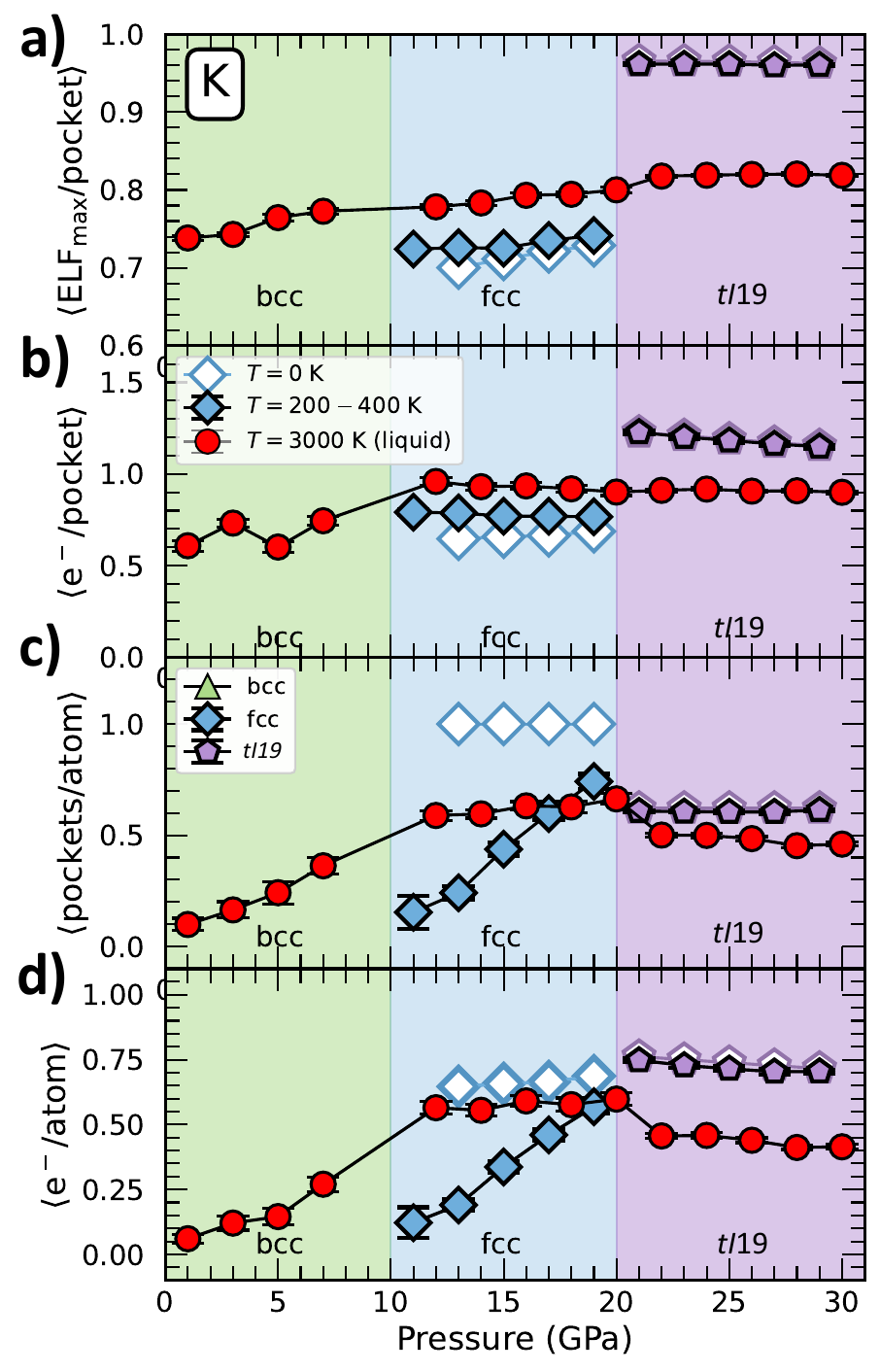}\\
    \includegraphics[width=6.7cm]{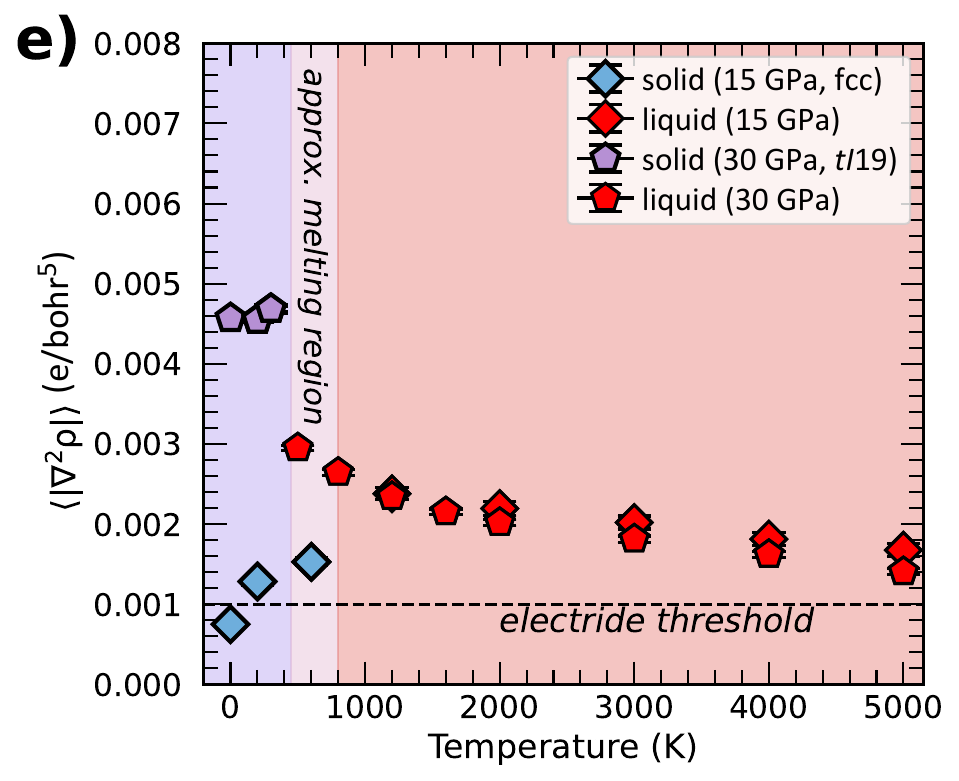}
    \caption{Potassium: The average a) ELF local maximum value of each pocket, b) number of electrons per pocket, c) number of pockets per atom, and d) number of electron pockets per atom vs pressure for K. The different shading in plots a-d represents the range of stability of the different phases at 0~K. The empty symbols represent $T = 0$~K data, while the filled symbols represent data at $T>0$. Red circles correspond to liquid K. The standard error bars for most data points are smaller than the markers.
    e) The average absolute value of $\nabla^2 \rho(\mathbf{r}_0)$ as a function of temperature for K fcc at $ \sim$15~GPa (blue diamonds) and K \emph{tI}19 at $\sim$ 30~GPa (purple hexagons). The approximate melting region encompasses the melting points of both structures.}
    \label{fig:K_e-_per_pocket}
\end{figure}

We heated K bcc at 200 K, K fcc at 400 K, and K \emph{tI}19 at 300K. K bcc did not have ELF values $\geq$ 0.7 at 250 K, just like at 0 K. Upon melting, K liquid had less than 0.9 electrons per pocket below 10 GPa, which corresponds to the pressure range that K bcc is stable in. As a result, K liquid 10 GPa is not an electride. For K fcc and \emph{tI}19, we found ELF local maxima $\geq $ 0.7 for the solids and when melted. K fcc, however, was still below the 0.9-electron per pocket threshold, even when heated (Fig. \ref{fig:K_e-_per_pocket}a).
K \emph{tI}19 is the only solid K electride we simulated, and we see that it is the only K solid structure with more than 0.9 electrons per pocket across temperatures. 
While there is less charge per pocket when K \emph{tI}19 melts, the liquid K at that pressure range was roughly one electron per pocket. We see an increase in ELF values starting at 13 GPa onward. According to Zong \emph{et al.}~\cite{zong_free_2021}, there is a liquid-liquid transition from the free electron to the electride liquid starting at 10 GPa, before fully becoming an electride liquid at 20 GPa. \revisiontwo{We also see a jump in the number of electrons per pocket in this pressure range, which shows that our methods are consistent with the existing literature.}

We also plotted the Laplacian with respect to temperature for K \emph{tI}19 at 30 GPa and K fcc at 15 GPa (Fig.~\ref{fig:K_e-_per_pocket}). Since K fcc is not an electride, the charge density Laplacian was lower than that of the liquid at $\sim$ 15 GPa. Even though the Laplacian value of K fcc is above our proposed threshold, it still has less than 0.9 electrons per pocket. Unlike Na, there was a gradual decrease in the charge density Laplacian upon melting for K \emph{tI}19, indicating that there was less of a change in electride behavior with increasing temperature. K \emph{tI}19 does not have a band gap as a solid electride \cite{woolman_structural_2018}, so K still has some metallic behavior as both a solid and a liquid. As a result, there is no major shift upon melting like that of Na hP4.

\subsubsection{Magnesium}

We investigated the effect of temperature on the electride behavior for Mg by thermalizing the bcc, fcc, sh, and sc structures at 4,000~K. We also studied the electride properties of liquid Mg along the 20,000 K isotherm to ensure that it remained liquid at all pressures (see Table \ref{tab:MD_table}).
We found that the Mg structures exhibit strong localization at 0~K with some charge pockets containing up to 2 electrons. 
Out of these structures, Mg bcc is the only one that does not seem to be an electride at 0 K, which agrees with previous studies~\cite{gorman_experimental_2022}.

\begin{figure}[!hbt]
    \centering
    \includegraphics[height=1.5cm]{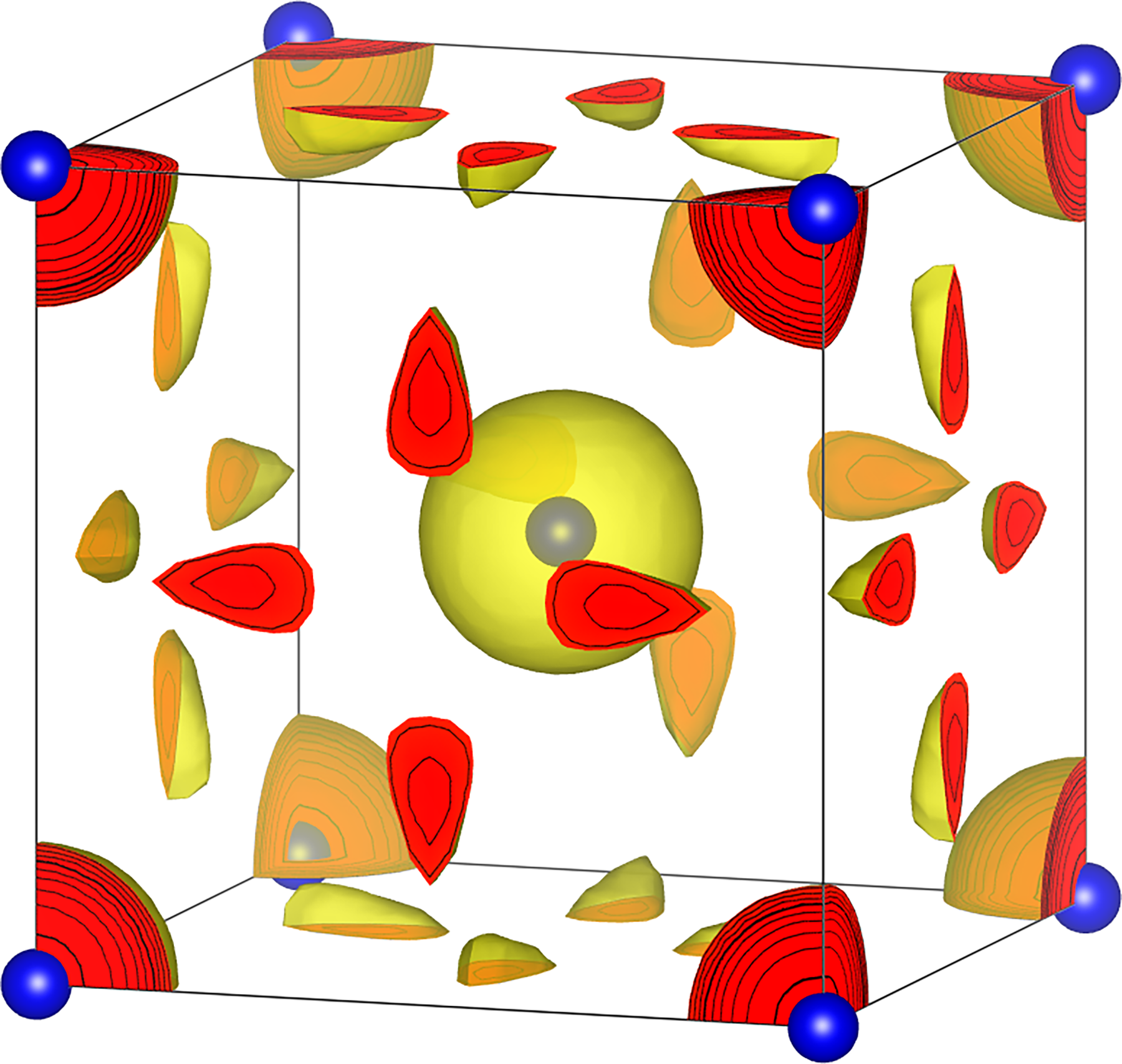}
    \includegraphics[height=1.5cm]{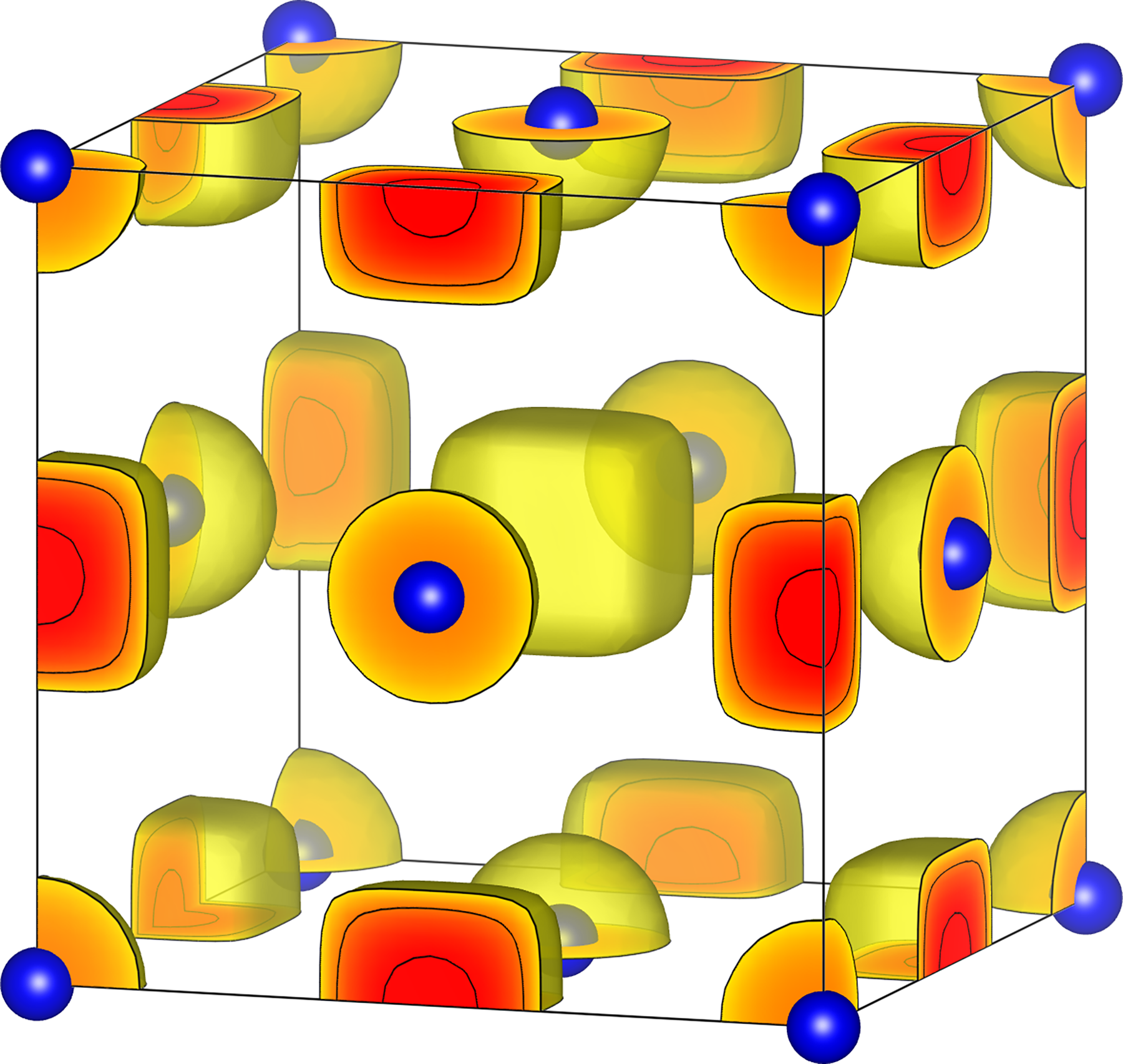}
    \includegraphics[height=1.5cm]{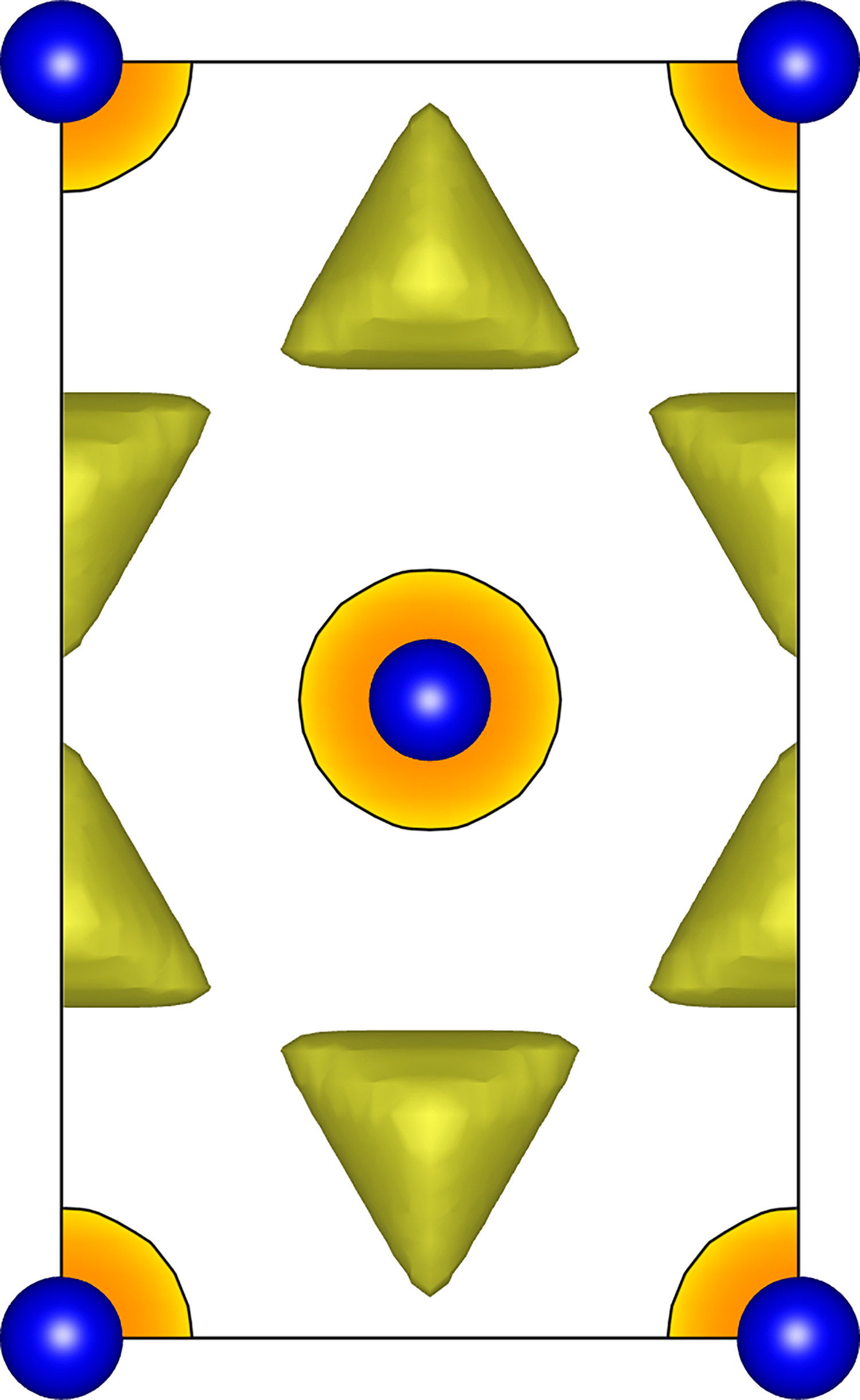}
    \includegraphics[height=1.5cm]{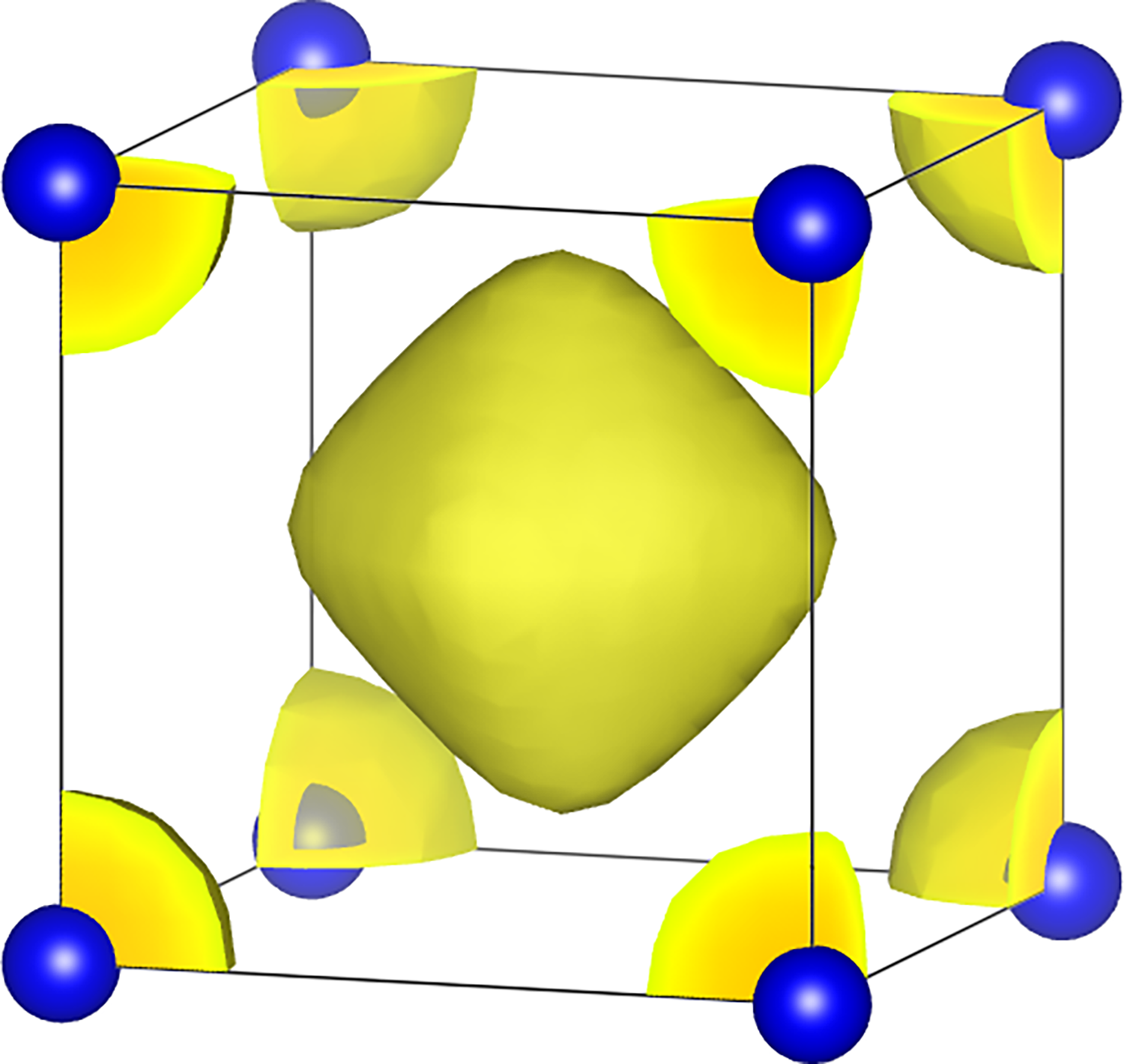}
    \hspace{0.1cm} \\
    \includegraphics[width=7.5cm]{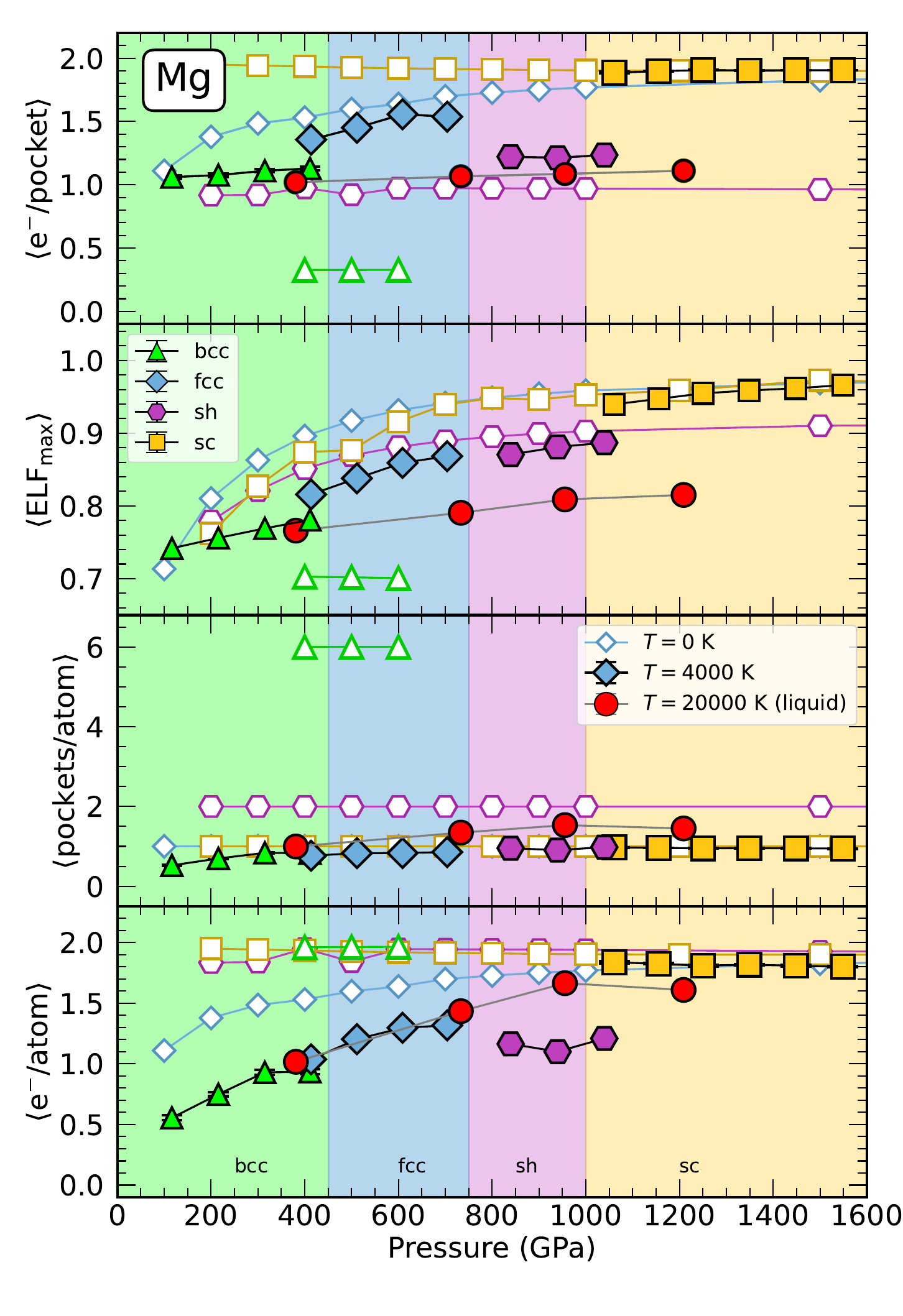}
    \caption{Magnesium: (from top to bottom) The average number of electrons per pocket, the average ELF local maximum value at the center of each pocket, the average number of pockets per atom, and the average number of electron pockets per atom vs pressure for Mg. The different shading represents the range of stability of the different phases at 0~K. The empty symbols represent $T = 0$~K data, while the filled symbols represent data at $T>0$. Red circles correspond to liquid Mg. The standard error bars for most data points are smaller than the markers.} 
    \label{fig:Mg_e-_per_pocket}
\end{figure}

In Fig.~\ref{fig:Mg_e-_per_pocket}, we show the average number of electrons per ELF pocket (integrated charge enclosed by the ELF isosurface) for each of the solid Mg structures of Mg as a function of pressure along the 0~K and 4000~K isotherms and liquid Mg at 20,000~K, together with the ELF value at the center of the pockets (local maximum), number of ELF pockets per atom, and the total integrated charge per atom over all pockets.
At pressures below 300~GPa, the small pockets formed in the bcc structure, which have a similar topology as that of bcc-Si (see Fig.~\ref{fig:Si_stats_vs_pressure}), do not meet the ELF value threshold criterion (ELF $\geq$ 0.7) at 0~K, so we set the enclosed charge to zero. However, at 4000~K the atomic vibrations open larger volumes in the interstitial sites of the bcc lattice, which allows more space for electrons to localize. The high temperature generates a sudden increase in the number of pockets per atom at lower pressures and, on average, there are just 0.5-0.8 pockets per atom at 4000~K that can accommodate a charge of 1e in each. This means that each Mg atom contributes 0.5-1 electrons localized in all pockets when the bcc crystal is at 4000~K.

\begin{figure}[!hbt]
    \centering
        \includegraphics[width=3.5cm]{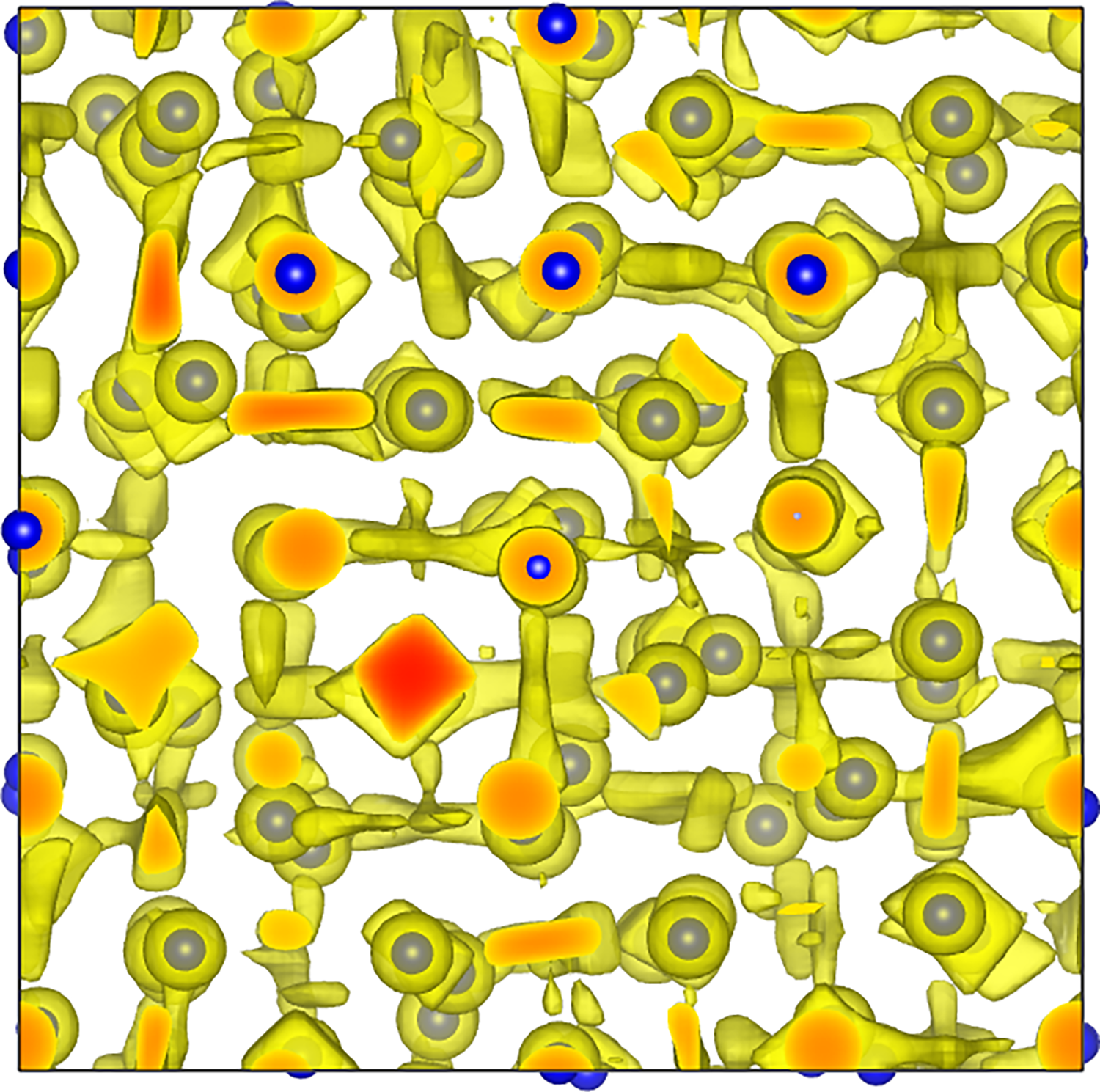}
        \includegraphics[width=3.5cm]{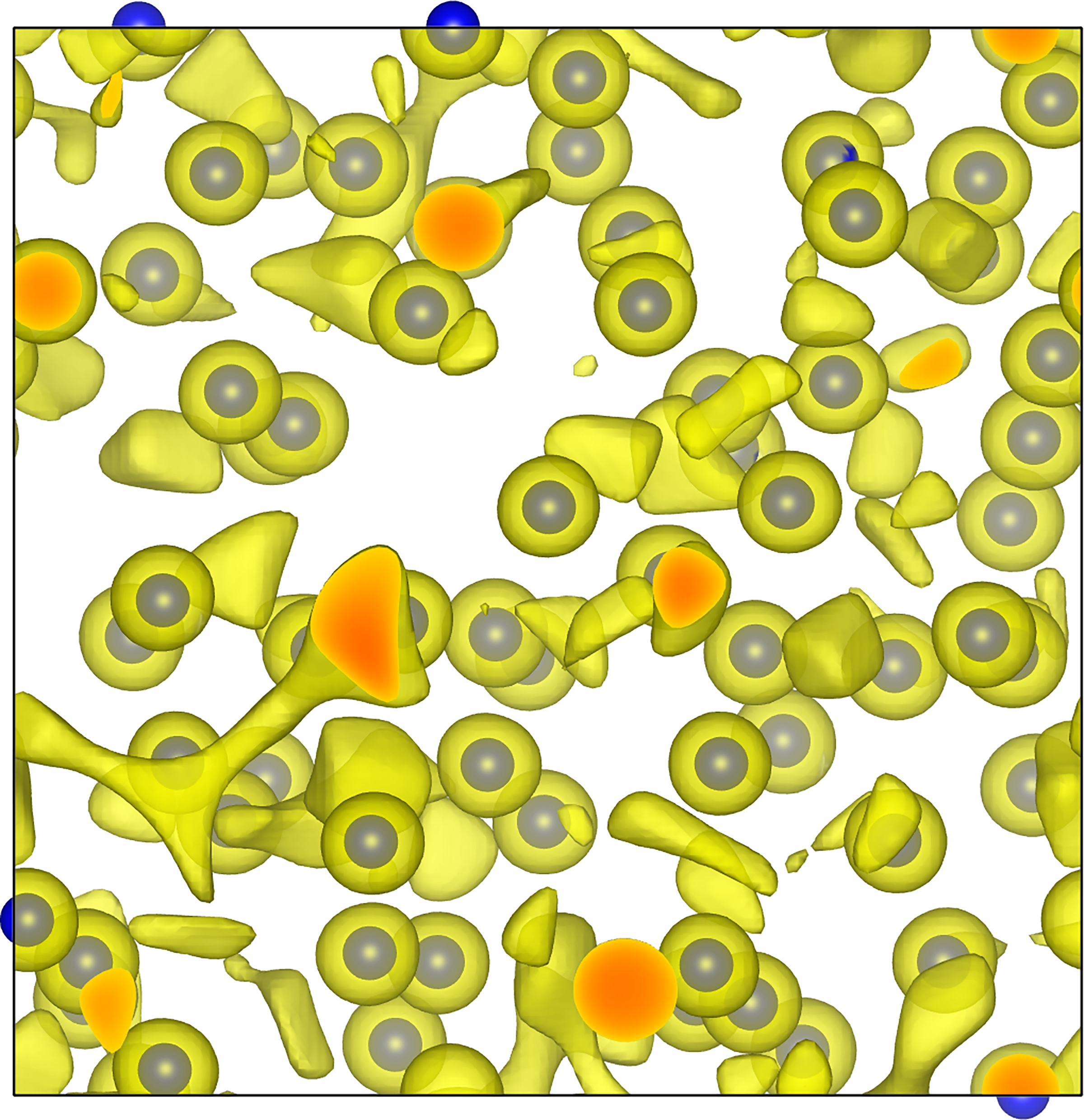} \\
    \includegraphics[width=7cm]{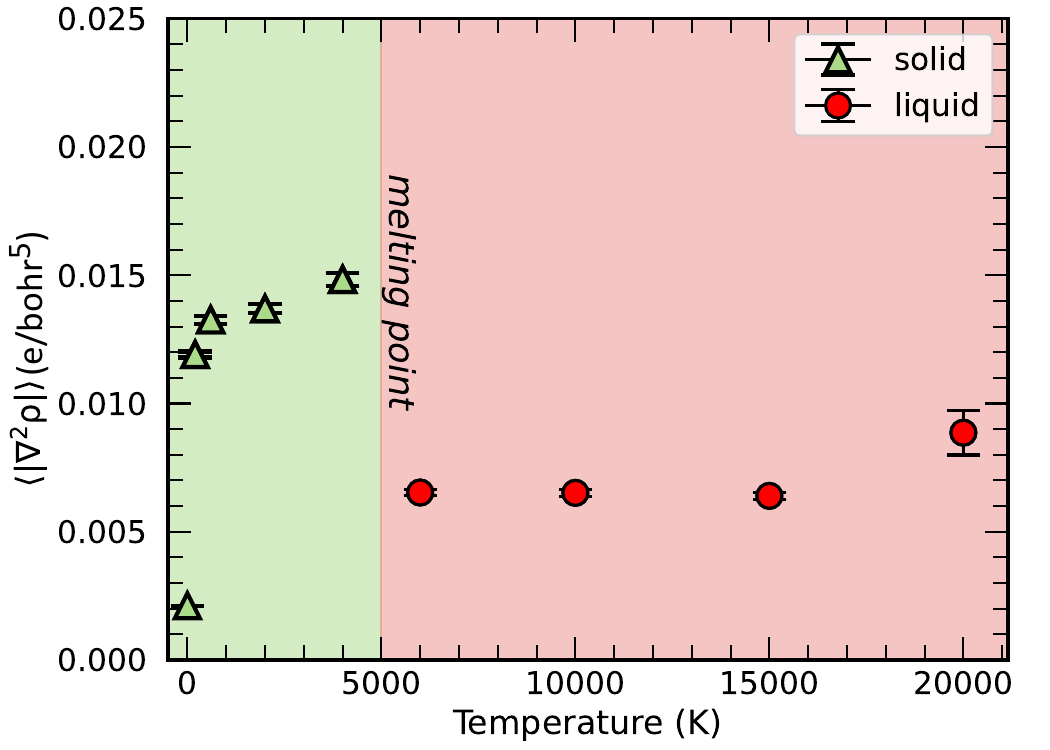}
    \caption{ELF isosurface (yellow, isovalue = 0.7) for solid, bcc Mg at 4000 K (top left), and liquid Mg at 20,000 K (top right). The panel below shows the average absolute value of the charge density Laplacian as a function of temperature for Mg along an isochore of density = 5.81 g~$\rm{cm^{-3}}$ ($\sim$400 GPa), starting from bcc Mg at 0 K. The melting point is estimated to be around $T = 5000$ K at this pressure.
    }
    \label{fig:Mg_BCC}
\end{figure}

Beyond 300~GPa, the electron localization in bcc becomes strong enough at 0~K to form 6 pockets per atom (12 per unit cell with 2 atoms, see Fig.~\ref{fig:Mg_BCC}), but they contain a charge that is less than 0.5 electrons each. Even though the total amount of localized electrons per atom is close to 2 electrons, the sparse distribution of charge in the ELF basins makes Mg bcc at 0 K not an electride.
When heating Mg bcc from 0 K to 4000 K at 400 GPa, there is a jump from 0.3 to 1.09 electrons per pocket (Table~3). There is also a jump in the charge density Laplacian from  $-1.7\times 10^{-3} $ to $-1.2\times 10 ^{-2}$  $e/\mathrm{bohr}^5$ (Fig.~\ref{fig:Mg_BCC}). This jump in both charge and Laplacian can be a sign that Mg bcc is ``transitioning" into an electride when heated. Mg fcc is the next stable phase when increasing pressure, and since the Mg bcc structures lose some of their symmetry when heated, that can make room for the charge pockets to form. Upon melting, we see the charge density Laplacian drop similar to liquid K (Fig. \ref{fig:K_e-_per_pocket}).

\begin{figure}[!hbt]
    \centering
    \includegraphics[width=3.6cm]{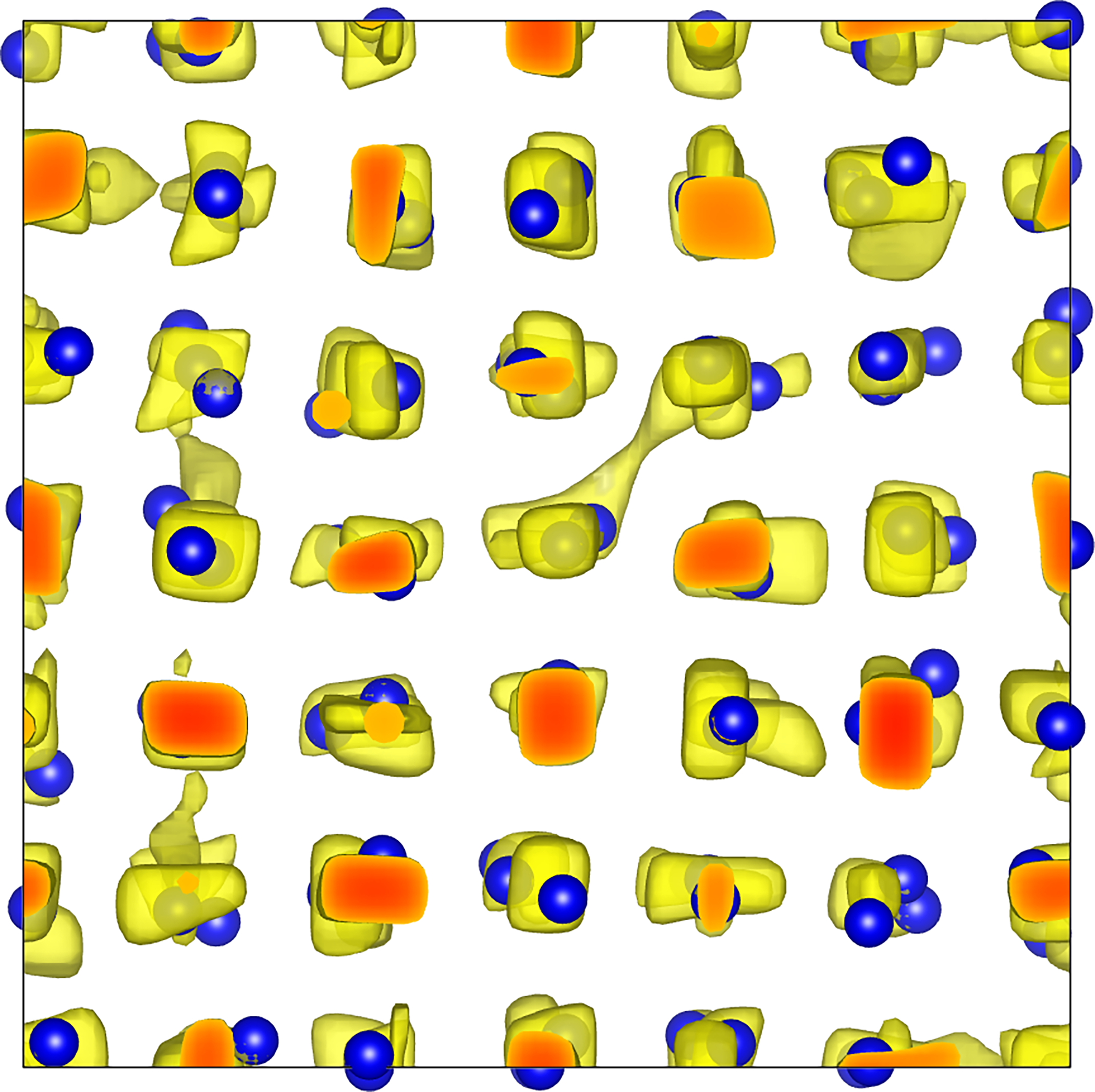} \hskip5mm
    \includegraphics[width=3.7cm]{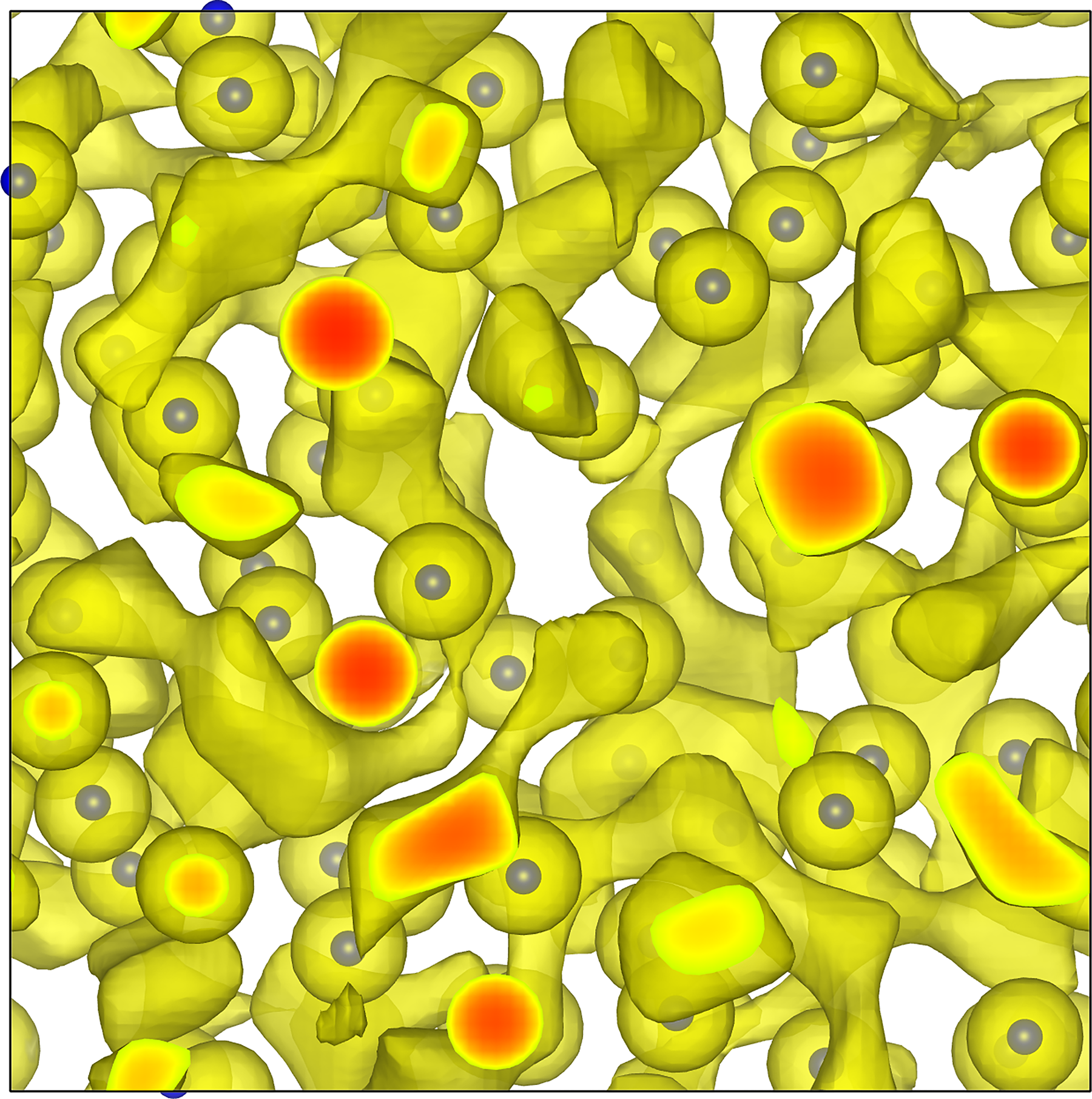} \\
    \includegraphics[width=3.6cm]{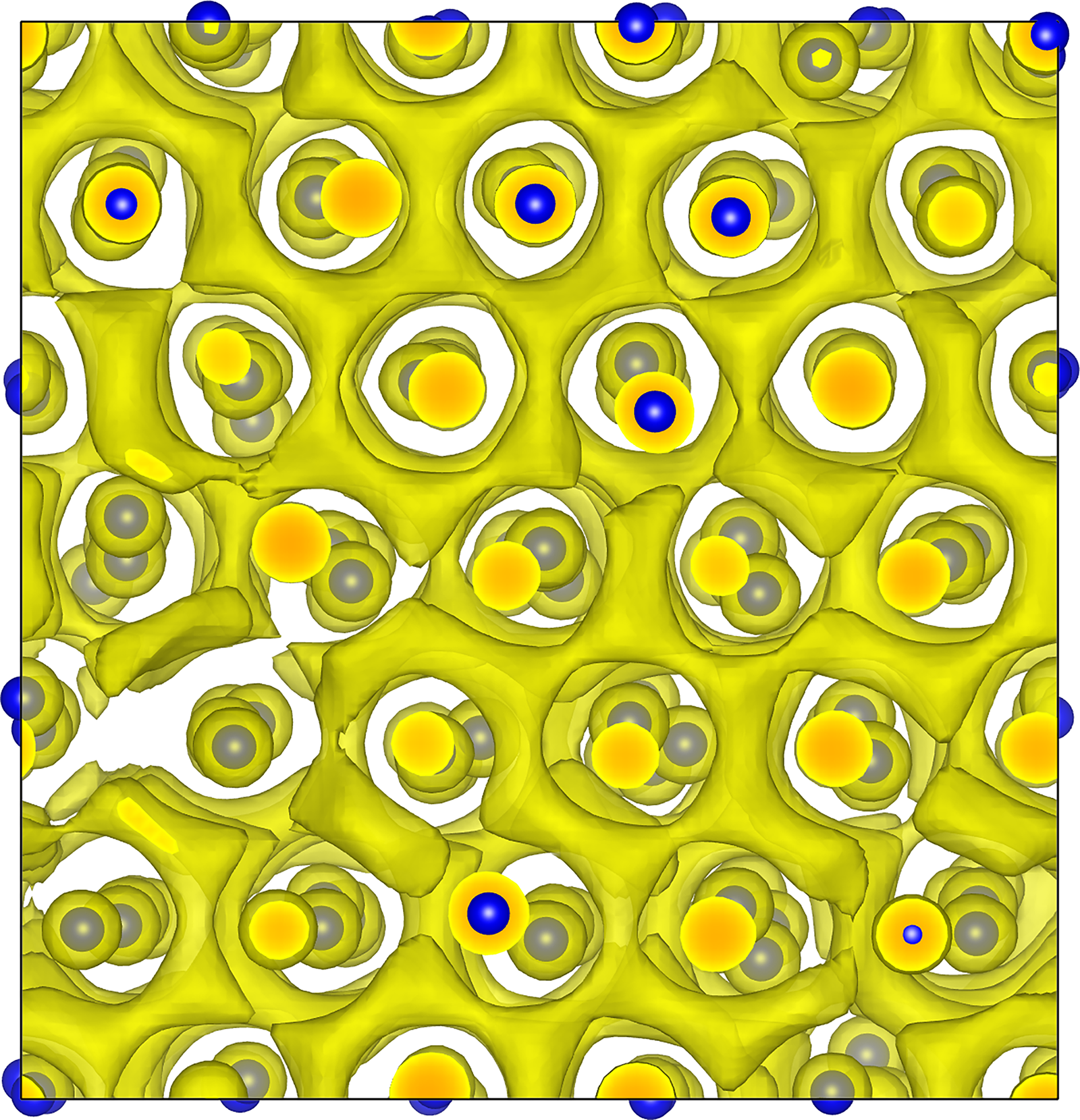}     \hskip5mm   
    \includegraphics[width=3.7cm]{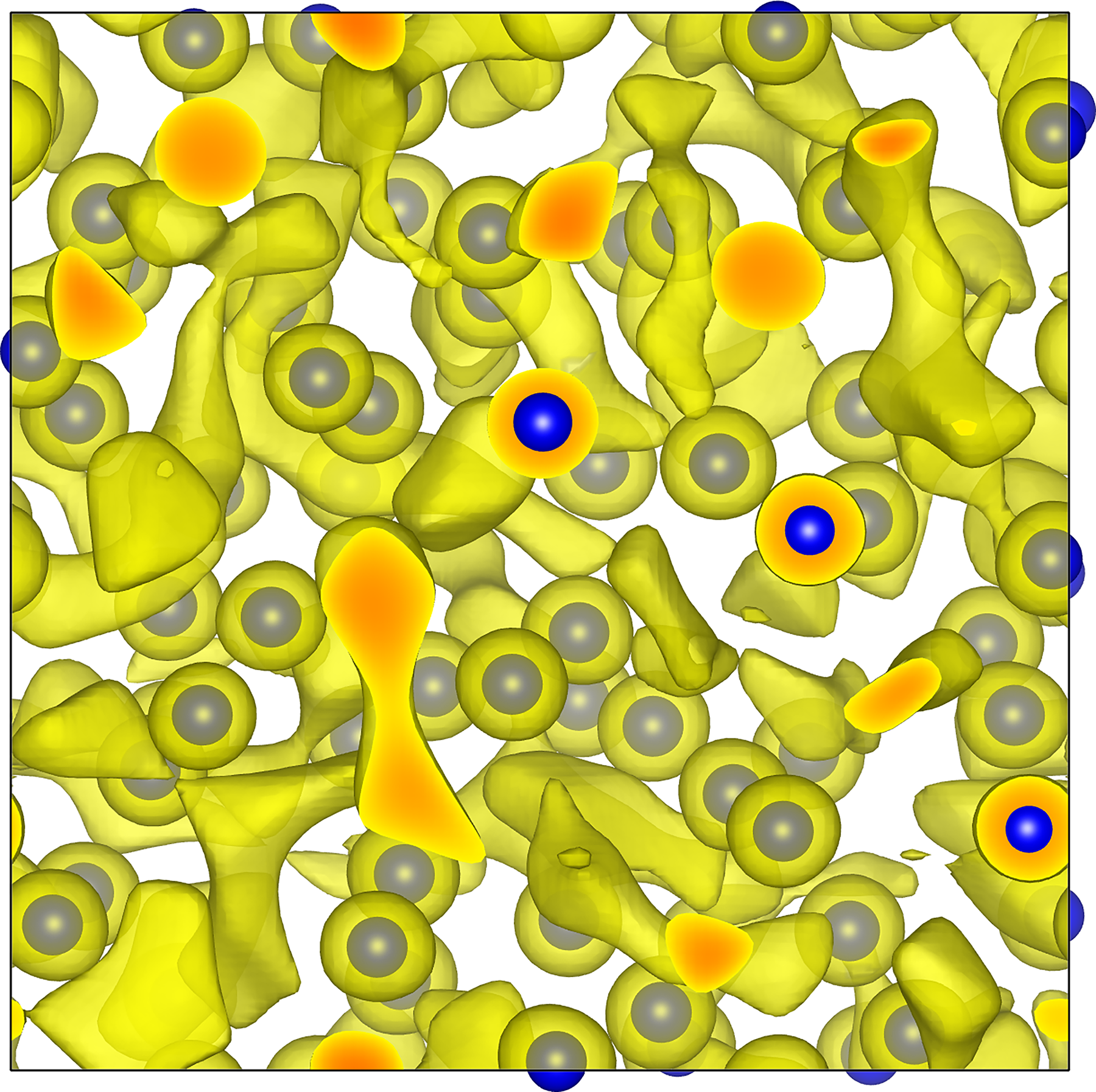}\\
    \includegraphics[width=3.6cm]{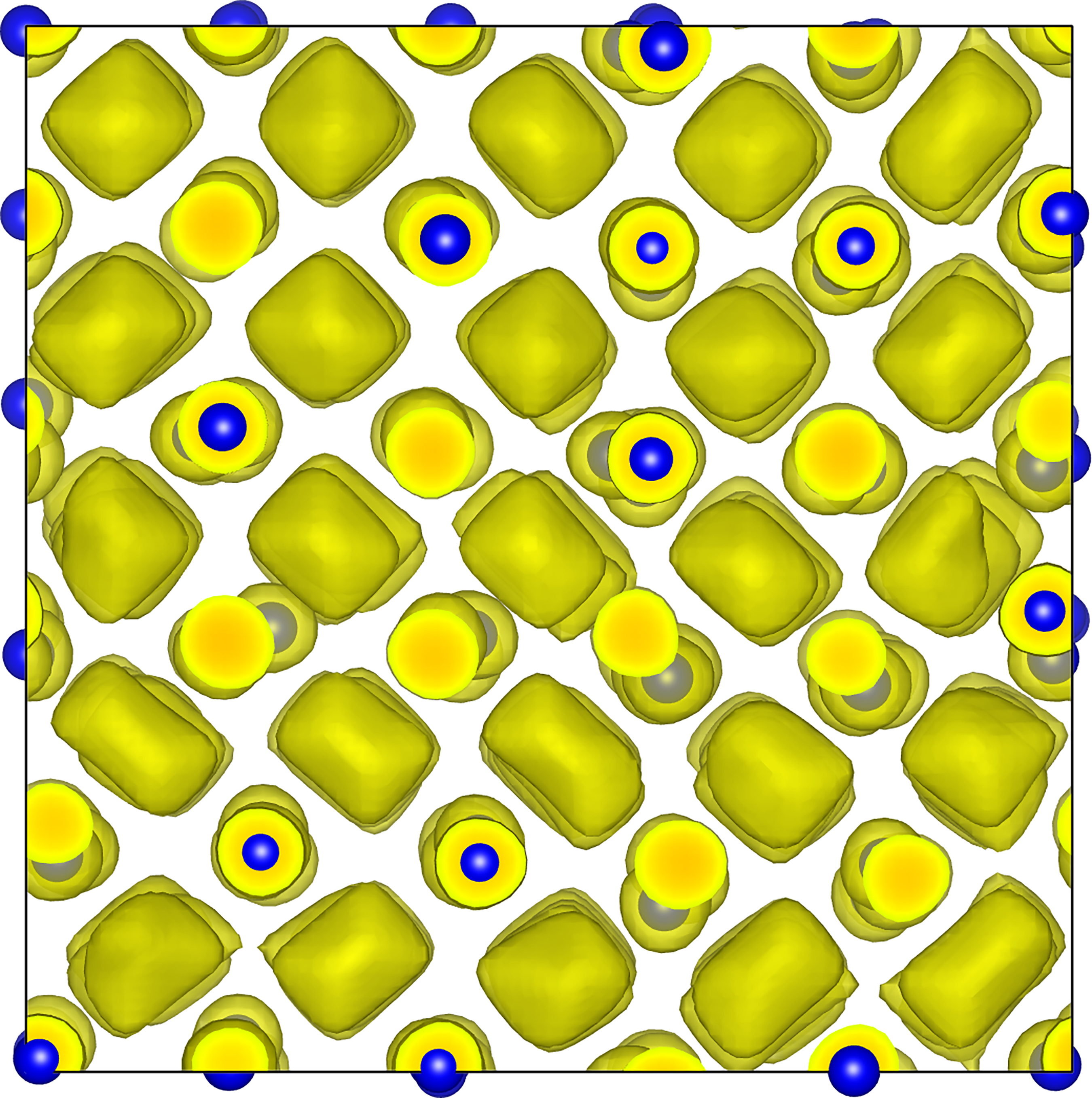} \hskip5mm
    \includegraphics[width=3.7cm]{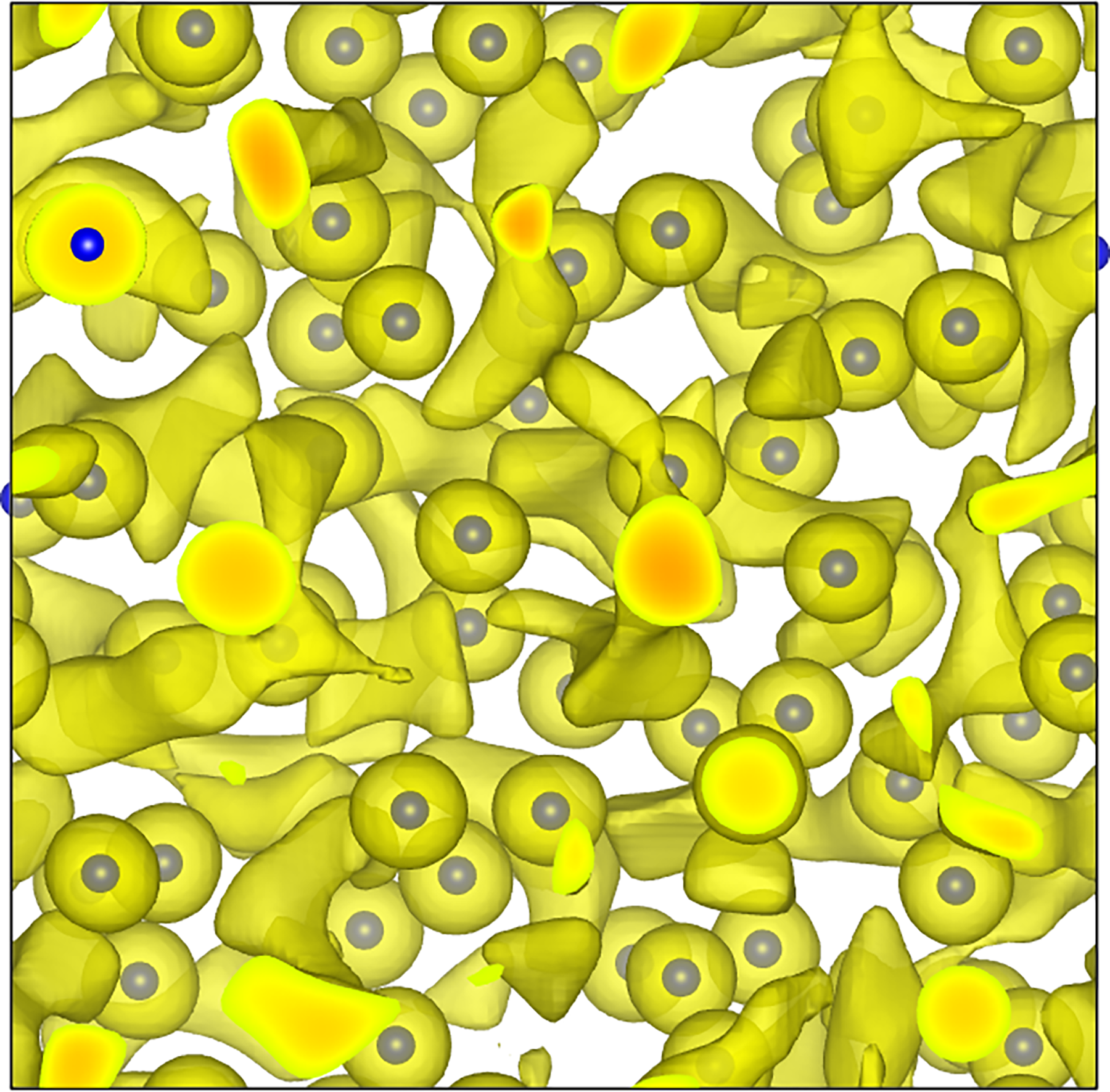}
    \caption{ELF plots for magnesium in three structures: fcc (top row), sh (middle row), and sc (bottom row). The first column shows electron localization in the unit cells at 4000~K. The second column shows snapshots of the liquid phase at 20,000 K. The pressures are 700~GPa for fcc, 900~GPa for sh, and 1300~GPa for sc. }
    \label{fig:Mg_FCC_SH_SC}
\end{figure}

Upon melting, the number of electrons per pocket remained the same as the high-temperature bcc structure. The liquid has a very similar behavior to the high-temperature bcc solid, with a very similar number of electron pockets per atom.
Based on our proposed criterion, Mg bcc is close to an electride when heated, and the liquid is clearly an electride, as it satisfies our five criteria, in agreement with previous indications of electride behavior in liquid Mg~\cite{gonzalez-cataldo_structural_2023}. 
The other Mg structures (fcc, sh, and sc) exhibit a clear electride behavior that is independent of pressure and temperature, as we see in Fig.~\ref{fig:Mg_e-_per_pocket}. The number of pockets per atom and integrated charge in these pockets is the same at 4000~K as at 0~K, which shows that the properties of these pockets remain stable at high temperature. The only exception is the sh phase, which exhibits a smaller number of ELF pockets per atom, that is less than half of what it exhibits at 0~K (about 0.94 pockets per atom). This results in a reduced number of total electrons per atom contained in pockets (1.2e instead of 2e).
The topology of the pockets, shown in Fig.~\ref{fig:Mg_FCC_SH_SC}, remains almost unaltered at high temperature for the solids and, remarkably, maintains almost the same properties in the liquid phase, such as the amount of pockets per atom and the charge contained in them, despite the lack of symmetry.

As we observed for Si, the electron localization is almost independent of pressure for a given structure, and we do not observe major changes unless there is a structural transition or the material melts. While pressure slightly increases ELF value at the local maximum, the amount of charge integrated in the ELF basins is basically the same at all pressures. It is important, however, to note that having localized electrons is necessary but not sufficient to classify a material as an electride alone. There must be enough localized charge present in the ELF basins to conclude that the electron localization is strong; thus, our five proposed criteria and threshold for both localization and integrated charge. 

\subsection{XRD}
\begin{figure}[!hbt]
    \centering
    \includegraphics[width=\linewidth]{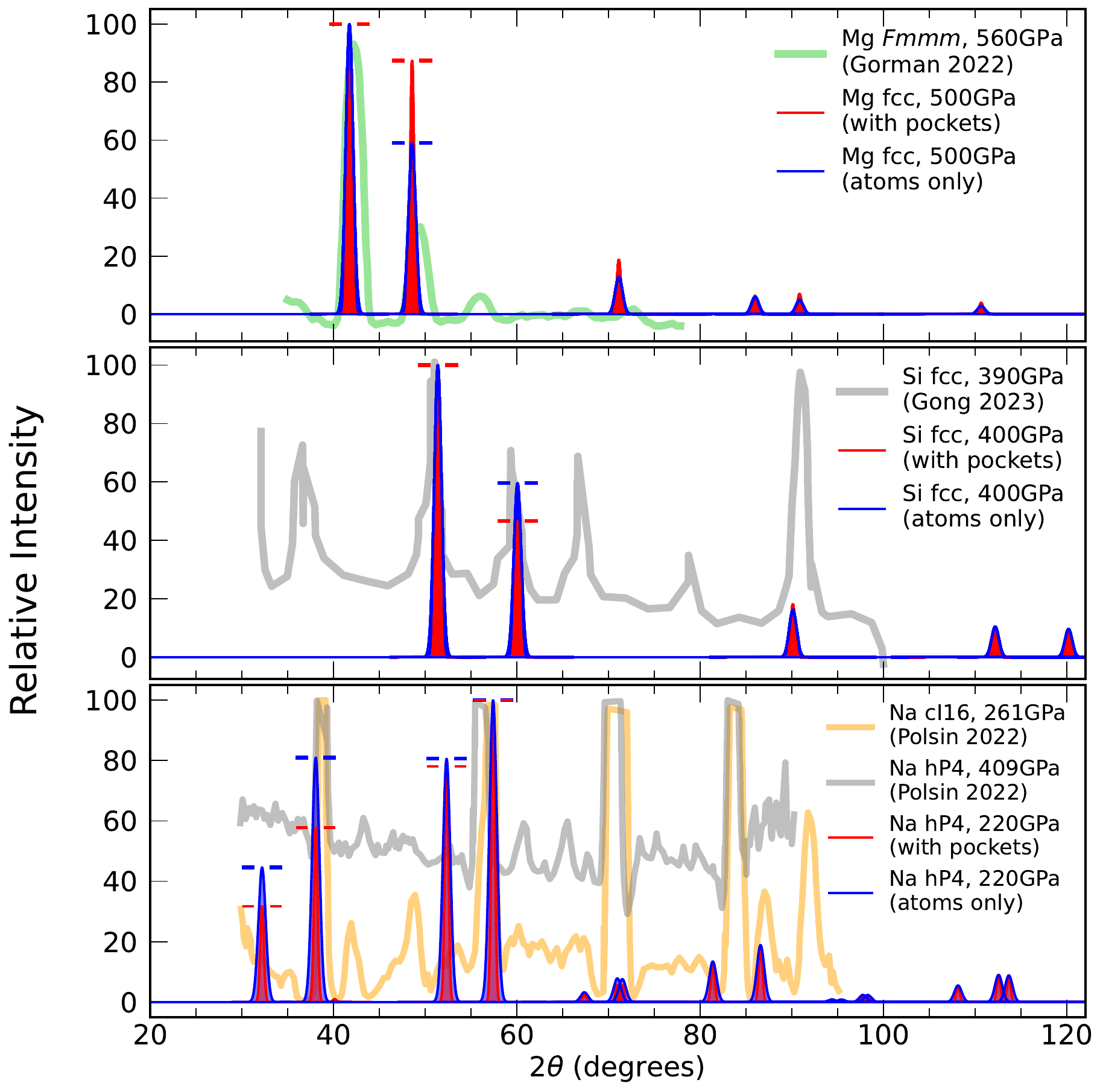}
    \caption{Predicted and measured XRD patterns for Mg fcc, Si fcc, and Na hP4 structures. The pressures are 
    560, 400 GPa, and 220 GPa, respectively. The blue curve corresponds to diffraction patterns generated by atoms only, while the red curve include scattering from the localized electron pockets. The vertical axis is the relative intensities, where the largest peak is set to 100.
    Experimental XRD patterns of Mg come from Gorman \emph{et al.} \cite{gorman_experimental_2022}, from Gong \emph{et al.} \cite{gong_x-ray_2023} for Si (at 390 GPa), and from Polsin \emph{et al.} \cite{polsin_structural_2022} for Na (at 409 GPa). The Mg $Fmmm$ structure from Gorman \emph{et al.} \cite{gorman_experimental_2022}, which is a slightly distorted fcc cell. The additional peaks in Si fcc structure that do not match our predicted pattern come from the pinhole material in the experiments by Gong \emph{et al.} \cite{gong_x-ray_2023}.
    }
    \label{fig:XRD}
\end{figure}

Since Si in the fcc structure had 2 electrons in the charge pockets, we calculated the form factor of the pockets to predict the XRD pattern for wavelength of $\lambda = 1.48$ \AA~ (Fig.~\ref{fig:XRD}, Table~\ref{tab:full_XRD_table}). This was done by taking the charge density grid of a single charge pocket and fitting a Gaussian along each axis. 
Since Si fcc is a cubic structure, the charge pocket was symmetric along all 3 directions. 
When compared to the Si fcc structure without the charge pockets, the amplitudes of the (200) XRD peak decreases by 12\% while the amplitude of the (220) peak increases by 2\%. The charge pockets in the fcc structure are not in the same plane as the atoms, which explains the reduction of the (200) amplitude. 

\begin{table}[!hbt]
    \centering
    \begin{tabular}{c c c c }
        \hline 
        2$\theta$ (deg.) & Peak & Intensity (\%) & Intensity (\%)\\
        & & without pockets & with pockets\\
        \hline
        Mg fcc\\
        41.7 & (111) & 100.0 & 100.0 \\
        48.6 & (200) & 59.0 & 82.3 \\ 
        71.1 & (220) & 12.9 & 17.3 \\ 
        86.0 & (311) & 6.0 &  6.5 \\ 
        90.8 & (222) & 4.9 & 6.4 \\\hline 
        Si fcc\\
        51.4 & (111) & 100.0 & 100.0 \\
        60.1 & (200) & 59.6 & 46.6\\
        90.1 & (220) & 16.2 & 18.2 \\
        112.2 & (311) & 10.4 & 10.4 \\
        120.2 & (222) & 9.7 & 9.1 \\\hline
        Na hP4\\
        32.2 & (010) & 44.6 & 31.8 \\
        38.0 & (011)  & 81.0 & 57.8\\
        52.3 & (012) & 80.6 & 78.0 \\
        57.4 & (110) & 100.0 & 100.0\\
        71.0 & (013)  & 7.6 & 5.6\\
        81.8 & (022) & 13.4 &  12.9 \\
        86.6 & (004) & 18.9  &  18.5 \\\hline
    \end{tabular}
    \caption{Predicted XRD patterns for Mg 
    fcc (560 GPa) phase, Si in the fcc phase (400 GPa), Na in the hP4 phase (220 GPa). The columns are 2$\theta$ values in degrees and the Miller indices of the x ray peaks, and their relative intensities before without and with consideration of the charge pockets.}
    \label{tab:full_XRD_table}
\end{table}

We also calculated the XRD patterns of Mg fcc structure with and without the charge pockets (Fig.~\ref{fig:XRD} and Table~\ref{tab:full_XRD_table}). The amplitude of the (200) peak increased by 22\% when the charge pockets were included. This is expected because the charge pockets are in the plane of the atom (Fig.~\ref{fig:FCC_Mg_vs_Si}). Conversely the charge pockets of the Si fcc structure do not lie in the same plane and thus cause the intensity of (200) peak intensity to decrease, as we have discussed for Mg above. 
For the Mg fcc structure, the amplitude second peak (011) in the calculated spectra is higher than seen in the experiments by Gorman \emph{et al.} \cite{gorman_experimental_2022}. We attribute this deviation to noise and interference from other materials in experiment.

We calculated the XRD patterns of Na hP4 structure at 220 GPa (Fig.~\ref{fig:XRD} and Table~\ref{tab:full_XRD_table}), a pressure that can be attained with diamond anvil cell. When accounting for charge pockets, we saw a 22\% decrease in the amplitude of the (011) peak, suggesting that the reflections from charge pockets destructively interfere with that of the atoms. We compare our simulated XRD pattern with Polsin \emph{et al.}'s experimental pattern. Some but not all peaks are found to be in good agreement.  

In summary, we find that the pockets do not contain sufficient charge to fundamentally change the XRD patterns but they can alter the amplitudes of the existing peaks. For example, in the Na hP4 structure (space group $P6_3/mmc$), the atoms occupy the Wyckoff positions $a$ and $d$, while the charge pockets occupy the Wyckoff position $c$. In the Si fcc structure, the atoms occupy Wyckoff positions $a$ while the charge pockets occupy position $c$. In both cases, they do not create new XRD peaks but alter the amplitude of existing ones.



\section{Conclusions}

We first showed with static DFT calculations that silicon becomes an electride in its fcc structure at 400 GPa, but then loses its electride features in the bcc phase at yet higher pressure. Then we performed DFT-MD simulations of solid and liquid Na, K, Mg, and Si over a wide range of pressures and temperatures. We found that all structures that exhibited electride behavior in the ground state remained electrides when the solids were substantially heated and when the structures were melted. The disorder in nuclear position did not fundamentally alter the electronic properties but may affect them in two ways. 1) When fcc Si is heated, we determined that the average charge per pocket and the number of charge pockets decrease slightly. 2) When materials in the bcc structures were heated, e.g., the thermal disorder of the nuclei caused some nearby charge pockets to merge, increasing their average charge while the total amount of charge of all pockets combined was found to be slightly less than that in the ground state. 

Then we compared the electride structures across materials, phases, and pressure-temperature conditions. Based on the qualitative criteria by Postils \textit{et al.}~\cite{Postils_molecular_electride_2015}, we developed the following quantitative criteria for to classify high-pressure electrides: 1) They contain pockets of electronic charge at interstitial sites where the electronic localization function (ELF) exceeds the value of 0.7. 2) The charge pockets contain at least 0.9 electrons. 3) Finally, we require that the Laplacian of the charge density be negative and its magnitude to be greater than $10^{-3}\ e/\mathrm{bohr}^5$. According to these criteria, Si fcc, Na \emph{cI}16, Na hP4, K \emph{tI}19, Mg fcc, Mg sh, and Mg sc are electride materials at T=0~K, while Si bcc, Mg bcc, K bcc, and K fcc are not. One should note that there are examples in the literature where electride behavior is associated with smaller ELF values such as 0.45 for C12A7:e- \cite{C12_matsuishi_2003,C12_Li_2004}. These ambient-pressure electrides are often synthesized by altering the stoichiometry of the material to make them electron-rich. As a result, these excess electrons are delocalized in order to stabilize the molecular or crystal structure. Furthermore, many ambient-pressure electrides, such as Ca$_3$Hf$_2$Pb$_3$~\cite{Li_Ca5Pb3_2021} and Cs$^+$(18C6)$_2$e$^-$~\cite{dawes_first_1986}, rely on having excess electrons to achieve the electride state. This mechanism differs from that of high-pressure electrides that we characterized in this article, where electrons in the highest valence states are destabilized by interactions with neighboring atoms. 

We conclude by predicting the pattern of XRD measurements, which we consider to be the primary laboratory technique to directly identify electride behavior of materials. For Na and Mg, we predict that the charge pockets alter the XRD intensities by approximately 20\%. If XRD measurements at high pressure could be improved to determine peak intensities to that level of accuracy, it would offer experimental pathways to verify the robust theoretical criteria that we have put forth for the electride classification of material in this article. 

\section{Supporting Information}
Additional details on equations of state, DFT setup, ELF visualizations, non-nuclear charge statistics, radial distribution functions, and charge density critical points.

\begin{acknowledgements}
We acknowledge beneficial discussions with Stefano Racioppi and Reetam Paul. S.A. was supported by the National Science Foundation Graduate Research Fellowship Program DGE 2146752. All authors received funds from the Department of Energy grant DE-NA0004147.
\end{acknowledgements}


\end{document}